\documentclass[onecolumn]{aastex631}
\usepackage{graphicx}
\usepackage{amsmath}
\usepackage{amssymb}
\usepackage{natbib}

\shorttitle{Solar activity index comparison}
\shortauthors{Zharkova\& et al}

\begin{document}
 
\title{Comparison of solar activity proxies: eigen vectors versus averaged sunspot numbers }

\author{Valentina V. Zharkova}
\affil{Department of MPEE, University of Northumbria, Newcastle upon Tyne,  UK}
\affil{ZVS Research Enterprise Ltd., London, UK} 
\email{valentina.zharkova@northumbria.ac.uk}
\author{Irina Vasilieva}
\affil{Department of Solar Physics, Main Astronomical Observatory, Kyiv,  Ukraine}
\affil{ZVS Research Enterprise Ltd., London,  UK}
\email{vasil@mao.kiev.ua}
 \author{Simon J. Shepherd}
\affil{PRIMAL Research Group, Sorbonne Universite, Paris,  France}
\affil{ZVS Research Enterprise Ltd., London, UK} 
\email{profsimonshepherd@gmail.com} 
\author{Elena Popova}
\affil{Centro de Investigación en Astronomía, Universidad Bernardo O’Higgins, Santiago, Chile}
\email{popovaelp@hotmail.com}
 \begin{abstract}
We attempt to establish links between a summary curve, or modulus summary curve, MSC, of the solar background magnetic field (SBMF) derived from Principal Component Analysis, with the averaged sunspot numbers (SSN). The comparison of MSC with the whole set of SSN reveals rather close correspondence of cycle timings, duration and maxima times for the cycles 12- 24, 6,7 and -4,-3. Although, in 1720-1760 and 1830-1860 there are discrepancies in maximum amplitudes of the cycles, durations and shifts of the maximum times between MSC and SSN curves. The MSC curve reveals pretty regular cycles with double maxima (cycles 1-4),   triple maximum amplitude distributions for cycles 0 and 1 and for cycles -1 and -2  just before Maunder minimum. The MSC cycles in 1700-1750 reveal smaller maximal magnitudes in cycles -3 to 0 and in cycle 1-4 than the amplitudes of SSN, while cycles -2 to 0 have reversed maxima with minima with SSN. Close fitting of MSC or Bayesian models to the sunspot curve distorts the occurrences of either Maunder Minimum or/and modern grand solar minimum (2020-2053).  These discrepancies can be caused by poor observations and by difference in solar magnetic fields responsible for these proxies. The dynamo simulations of toroidal and poloidal magnetic field in the grand solar cycle (GSC) from 1650 until 2050 demonstrate the clear differences between their amplitude variations during the GSC.  The use of eigen vectors of  SBMF can provide additional information to that derived from SSN that can be useful for understanding solar activity.
 \end{abstract}

\keywords{Sun: magnetic field -- Sun: solar activity -- Sun: sunspots  -- Sun: solar dynamo}

\section{Introduction} \label{intro}

Solar activity is a fundamental  process of generation radiation, energetic particles and waves affecting  the Earth and other planets, the climate and human lives. The points how sunspot activity varies in time and how it is linked to magnetic activity are very important issues investigated by many researchers. For the past two centuries a solar cycle was defined through sunspot numbers and this solar activity index was also used  for prediction of the  future solar activity while testing mechanisms of the solar dynamo providing conversion and transport of solar magnetic fields from the solar interior to its surface. The dynamo models operate with poloidal and toroidal magnetic fields \citep{Parker55}, with the first one being  the solar background magnetic field (SBMF), and the second one being the magnetic field of magnetic loops in active regions, which are embedded into the solar surface, whose roots look like sunspots. 
 Many researchers of 18-10 centuries did not have this knowledge because neither magnetic field was discovered nor  measured on the Sun.  
 
For the past 400 years and even longer than 1000 years, if Chinese observations included, sunspots were observed with different level of regularity  In the middle of 19 century a few experienced observers discovered  that dark spots on the sun, sunspots, appear rather periodically approaching maxima and minima within every 11 years \citep{Schwabe1843, wolf1850a, wolf1850b}. Based on this periodicity, the first index of solar activity was expressed with monthly sunspot numbers, or Wolf's number W, averaged from many observatories \citep{wolf1850a, wolf1850b}. For nearly three hundred years  Relative Solar Sunspot Number (SSN) introduced by Rudolf Wolf (referred to as the Wolf Numbers, W) are still used (www.sidc.be/silso/home) as the solar activity index. 

Sunspots were actively studied across various cultures and geographies of the Earth from visual observations before the invention of a telescope and from telescopic monitoring of the Sun after; although, these observations were affected by observational gaps and unquantified uncertainties (see \citep{hoyt1998b, SoonYaskell2003, Arlt2008, Arlt2009, Ogurtsov2013, Clette2014,  Arlt2016, Zito2016, Tamazawa2017, Hayakawa2017a, Hayakawa2017b, Willamo2017, Chatzistergos2017, Neuhauser2018, Munoz-Jaramillo2019, ArltVaquero2020, Carrasco2020, Simpson2020, Carrasco2021a, Carrasco2021b, Vokhmyanin2021, Hayakawa2021a, Hayakawa2021b}. 

In addition to the standardised Wolf sunspot numbers (WSN hereafter)) or the international sunspot number (referred to as SSN), \citet{hoyt1998b, hoyt1998a} introduced the group sunspot number (GSN)  in order to repair a  deficiency in observing small sunspots and managed to make compatible the resulting Group Sunspot Number (GSN) with the SSN. Although, these two series shown serious disagreements for the interval before 1885; and the GSN were not maintained after 1998. 

Other authors tried to identify the solar activity nature with  underlying solar magnetism giving the full cycle  for every 22 years because the leading magnetic polarity of sunspots is shown to change every 11 years  \citep{Hathaway02} see \citep[see also][]{Livingston2012,  Nagovitsyn2012} and  
managed  to reduce noise levels in the standardized sunspot numbers by accounting for smaller and more sporadic individual spots \citep[see e.g.][]{Hathaway2013, Carrasco2018}.
 \citet{Willamo2017} using the singular spectrum analysis method  did a new reconstruction  of sunspot  numbers marking centennial variability of solar activity and the modern grand maximum occurring in the second half of the 20th century. .\citet{Willamo2017} evaluated a stability of some key solar observers against the composite series concluding the Royal Greenwich Observatory dataset to be 10$\%$ too low before and relatively stable since the 1890s.

 Recently, \citet[][hereafter SS16]{Svalgaard2016, Svalgaard2017b} revisited the issue and brought up-to-date the Group Number series \citep{hoyt1998b}. 
 \citet{Svalgaard2017b}  compared four reconstructions of the number of sunspot groups (‘active regions’) restored by different authors with different methods for the period since AD 1900 where the solar data are of sufficient quality. and find that  The found that all four methods yield the same GSN series and suggested to investigate  any specific reasons of these methods giving  not agreeable results for the intervals before 1900. Despite the results of this research  \citep{Svalgaard2016, Svalgaard2017b} were severely criticised on the procedural grounds \citep[e.g.][]{Lockwood2016b}, the authors managed to successfully dismount this criticism \citep{Svalgaard2017c}. 
 
  \citet{Chatzistergos2017} using a daisy chain process with backbone (BB) observers  performed the calibration of each individual observer with a probability distribution function (PDF) matrix constructed considering all daily values for the overlapping period with the BB. Using the final series extending back to 1739 and including the data from 314 observers, \citet{Chatzistergos2017} concluded  that the series suggests rather moderate activity during the 18th and 19th century, that is significantly lower than predicted by other recent reconstructions applying linear regressions.
   
   Although, what is truly remarkable that since 1900 when the underlying sunspot data are plentiful and of good quality, all these reconstructions agree within a few percent, regardless of methodology and of claims of being superior to all the others \citep{Cliver2016}. This means that in spite of the criticisms, all the methods are equally satisfactory, provided the data are good and in particular that the ‘standard’ or ‘reference’ observer records are stable and free of anomalies.  

The recent research \citep{usoskin2021} using the isotope $^{14}$C production rate determined from the measurements of radiocarbon content in tree rings, reconstructed solar activity for the period 971 – 1900 (85 individual cycles).  \citet{usoskin2021} applied a Monte-Carlo method to find the lengths and strengths of cycles outside grand minima, which were agreeable with those derived from the direct sunspot observations after 1750. The authors reported three potential solar (or other source) particle events  occurring in 994, 1052 and 1279 AD around the maximum phases of solar cycles. Although, the authors reported some shifts in sunspot maxima and noted that the individual solar cycles can hardly be reliably resolved in the abundances of $^{14}$C isotopes, which can reliably mark only  grand solar minima (GSMs). 

Recently,  \citet{zharkova12, zhar15} suggested to use an additional proxy of solar activity - the eigen vectors of the solar background magnetic field  (SBMF) obtained from  the Wilcox Solar Observatory low resolution synoptic magnetic maps. . By applying the principal component analysis (PCA) to the synoptic magnetic data for cycles 21-23  \citep{zharkova12} and recently for 21-24 \citep{Zharkova2022} the authors identified and confirmed a number of eigen values and eigen vectors  from the SBMF representing magnetic waves of the solar surface generated by different magnetic sources in the solar interior. 

The first pair of eigen vectors covered by the largest amount of the magnetic data by variance, or  principal components (PCs), reflects the primary waves of solar magnetic dynamo produced by the dipole magnetic sources \citep{zhar15}. The temporal features of the summary curve of these two PCs shown a remarkable resemblance to the sunspot index of solar activity (representing toroidal magnetic field of active regions) for cycles 21-23 \citep{zhar15} and cycles 21-24 \citep{Zharkova2022}. This correspondence occurs despite the PCs and sunspot indices represent very different entities of solar activity: poloidal magnetic field for PCs and toroidal magnetic field for sunspot numbers.  However, their similarity and periodicity allowed  \citet{zhar15} to suggest using this summary curve of the two PCs of SBMF as a new, or additional, solar activity proxy. 

The advantage of using  the summary curve as solar activity proxy instead of the averaged sunspot numbers is a usage of the eigen vectors of the solar (poloidal) magnetic field oscillations,  derived from the surface magnetic field measurements on the whole solar disk which are expressed then by mathematical formulae and, thus, can be extrapolated for any time. The ample magnetic  full disk data  significantly reduces the errors  in the magnetic wave definition and introduces an extra-parameter, a leading polarity of SBMF \citep{zhar15},  which is  shown to be in anti-phase with the magnetic polarity of leading sunspots \citep{zharkov08}.   
 
The summary curve, calculated backward 800 years to 1200  and forward to 3200, reveals the very distinct variations of the 11-year cycle amplitudes in every 350-400 years, or grand solar cycles (GSCs) separated by grand solar minima (GSMs) when the amplitudes of 11 year cycles become very small, similar to those reported in Maunder,  Wolf,  Oort and other grand solar minima. The timings of GSMs are defined by the interference (so called beating effect) of two magnetic dynamo waves with close but not equal frequencies  defined by the different velocities of meridional circulation \citep{zhar15}.  The summary curve has also shown the modern GSM  to occur in the cycles 25-27, or in 2020-2053 \citep{zhar15, zhar2020}. 
  
This finding was confirmed recently by the other researchers \citep{Kitiashvili2020, Obridko2021}who obtained the modern grand solar minimum  to occur in cycles 25-27. The authors used the WSO synoptic magnetic field data similar to those by \citet{zhar15} and compared zonal harmonics of SBMF in the past four cycles  with 3D dynamo models. In addition, \citet{velasco2021} obtained the similar GSM  in cycles 25-27 using only average sunspot number  observations and  applying the machine learning  (ML) algorithm.

There were also two other reconstructions of the solar cycles represented by sunspots  using Bayesian approach \citep{velasco2021, velasco2022}. In the first paper \citep{velasco2021} a Bayesian model  fitted to the  later cycles from 1850 towards the modern times shown the occurrence of the modern grand solar minimum \citep{velasco2021}, similar to that predicted by \citet{zhar15}. The later Bayesian fit of the whole series of sunspot cycles including those in 17 century \citep{velasco2022} reveals a good correspondence of model cycles to the solar activity cycles while not matching the cycle durations or times of their maxima. This is  highlighting the problems with accuracy of the solar activity curves expressed via sunspot numbers  in early years when there were little or none observations. 

These findings  raise significant interest to a comparison of the sunspot activity indices with the indices derived from Bayesian approach and with the solar activity index derived from eigen vectors of the solar background magnetic field \citep{zhar15} that is a motivation for the current study.

The overview of solar activity indices defined by sunspots  and uncertainties in their definition are presented in Section \ref{ss_index}, the recent restoration of the sunspot index with Bayesian method and their  associated definition problems is described in Section  \label{ss_bayes} while the new eigen vectors derived from  the SBMF synoptic maps and their  comparison of  the summary curve of PCs with  sunspot indices are described in Section \ref{ss_pca}, conclusions are drawn in section \ref{conclusion}.
\section{Solar activity indices: averaged sunspot numbers} \label{ss_index}
Most of our knowledge about sunspots in the 18th century relies on sunspot drawings by J.C. Staudach  \citep[digitised by ][]{Arlt2008}). The current averaged sunspot numbers are provided  by \citet{SILSO} from 1730 until present. 
\subsection{Historical sunspot data: collection and problems} \label{ss_hist}
Since 17 century observations of sunspots  were carried out in Paris, Marseille, Lilienthal, Gottingen, Dresden, Berlin, Graz, Munich, Edinburgh, Venice, Milan, and Prague. But none of the observers even thought about how to generalise the observations ast it was believed that the appearance of sunspots was an accidental phenomenon.

\begin{figure}
\includegraphics[scale=0.39]{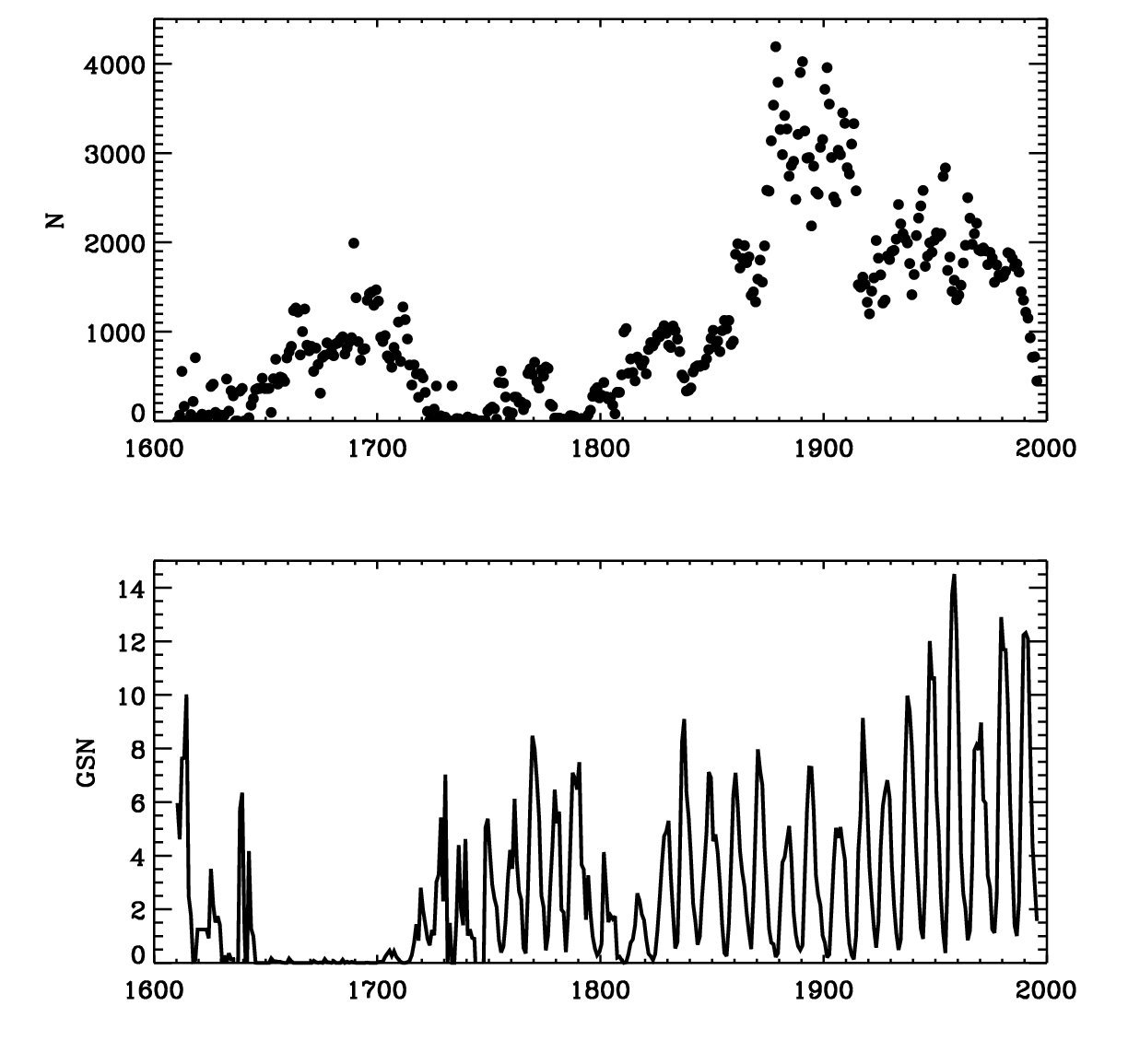}
 \caption{Top plot: the number of observations per year used to reconstruct the averaged sunspot numbers.  Bottom plot:  The averaged sunspot numbers reconstructed from 1600 until the present time \citep{SILSO}. }
  \label{obs}
\end{figure}

 In the second half of the 17th century to the beginning of the 18th century, sunspots were observed by Jan Hevelius in Danzig, Jean Picard in Paris, Martin Vogelius, and Heinrich Siverus in Hamburg, John Flamsteed in Greenwich, Georg Christoph Eymmart in Nuremberg, Philippe de la Hir in Paris, and some their other contemporaries \citep{Wolf1877, Usoskin2017}. For many years, the astronomers diligently took notes on the spots on the Sun. Gradually, the huge interest in observing sunspots, which existed soon after their discovery, was lost  and sunspot observations became scarce (see Fig. \ref{obs}, top plot).

From the middle of the 18th century, the number of observations has increased. In particular, Johann Kaspar Staudacher in Nuremberg and Christian Horrebu in Copenhagen made notes and sketches for 50 (more than 1000 observation days) and 15 (more than 3000 observation days) years, respectively \citep{Dreyer1903, Svalgaard2017}. At the same time, shorter series of observations were made by Ludovico Zucconi in Venice (almost 900 days of observations over 16 years) and Karl Schubert in Danzig (almost 500 days of observations over 4 years) \citep{Wolf1877}.

Information about solar activity in the period 1749-1799 is based primarily on data from only two observers: Christian Horrebow and Johann Caspar Staudacher \citep{hoyt1998b}. At the same time, a characteristic feature of solar cycles in the second half of the 18th century is a strong deviation from the average duration of the period of 131 $\pm$ 14 months \citep{Hathaway2015}  (cycles 3 and 4 lasted 9.0 and 9.3 years, respectively, and cycle 5 - 13.6 years). In this observation period, the various reconstructions of solar activity diverge significantly \citep{Karoff2019}.
At the end of the 18th century and the beginning of the 19th century, Pierre-Gilles-Antoine-Honore Flaugergue in Vivier, Augustine Stark in Augsburg, Johann Wilhelm Pastorff in Drossen, Frankvis Arago in Paris, and Johann Friedrich Julius Schmidt in Athens joined the Sun’s observers \citep{Wolf1877}. 

 \citet{Schwabe1843}  based on the records of daily observations between 1826 and 1843, discovered a frequency of occurrence of spots with a cycle to be approximately 10 years. Later Rudolf Wolff redefined the duration of a solar cycle finding  the maximum number of spots to be repeated every 11.1 years \citep{Wolf1852}. Wolf was the first researcher to organise permanent systematic observations of sunspots and introduced the concept of the daily “relative” sunspot number (Wolf sunspot number, WSN), international sunspot number, Zurich number): $W=k(f+10g)$,
where $g$ is the number of groups of spots, $f$  is the number of individual spots, and $k$ is the weight factor for the observer.  

\citet{Wolf1877} introduced a system WSN where the number of sunspots per day was determined by the main (preferred) observer. If the main observer was unable to count, then the definition from the secondary or tertiary observer with different weights was used. For each observer the individual observations are used to compute, first, monthly averages, then yearly averages  from the averages of all months, if in this year at least one observation is carried out. This algorithm allowed him to minimise the effect of clustering in time, and to introduces a noise stemming from the months with a few observations. 

A primary observer was selected based  on length of the observational series (as long as possible) and on the perceived “quality” of sunspot observations, e.g. suitable telescope, regularity of observations and a lack of detected problems. From 1849 to 1893, Wolf himself was the main observer, then the others were: Alfred Wolfer (Zurich) from 1894 to 1926, William Otto Brunner (Zurich) 1926–1944, Max Waldmeier (Arosa) 1945–1979. In the current times the International Sunspot Number has been provided since 1981 by the Royal Observatory of Belgium with Sergio Cortesi (Locarno) as the main observer. 

Wolf expanded the records 100 years ago, using Johann Kaspar Staudacher (Nuremberg) as the main observer from 1749 to 1787, Honore Flogerga (Vivier) from 1788 to 1825, and Samuel Heinrich Schwabe (Dessau) from 1826 to 1847. The scientists made very detailed sketches of the structure of sunspots, which are not lower quality in detail to even the best images taken with modern telescopes. 
Wolfer  joined Wolf in counting sunspots with a new telescope in 1877. As result, in the early 1880s, the observer factors for Wolf and other observers were altered, and essentially they started a new time series. This step caused a discontinuity in Wolf numbers in 1879 – 1883, which was unnoticed. 

 \subsection{Recent restorations of the sunspot data} \label{ssn}
A significant step in improving the sunspot series was made in 1998 by \citet{hoyt1998a, hoyt1998b} who published a revised sunspot series with sunspot groups (GSN) from 1610 based on the analysis of 455242 records of 463 observers. HS98 Used a special “fill-in” procedure to fill sunspot numbers for days with no observations, in order to “bridge” the data gaps. 

A number of authors tried to understand the differences between the group numbers (GSN) and Wolf’s sunspot numbers (WSN)  \citep{Hoyt1994, Clette2014, velasco2021}. 
The GSN series of \citet{hoyt1998b} is found to be more consistent and homogeneous with Schwabe’s data throughout the entire studied period  as found by \citet{Leussu2013} while the WSN records decreased by roughly 20$\%$ around 1848 because of the change of the primary observer from Schwabe to Wolf. 
Although,  the GSN reconstruction becomes very similar to Wolf’s reconstruction before the 1.25 correction factor was applied \citep{Hoyt1994}.

\citet{Clette2014} reported about a noticeable trend found and eliminated in the solar activity index derived from the observations of the Locarno Observatory, which was a reference observatory after 1980. Also \citet{Clette2014} derived the three-peak shape (so-called $\Psi$-type distribution) of the original GSN by \citet{hoyt1998a, hoyt1998b} for sunspot Cycle -1 with the peaks in 1736, 1739, and 1741.  Later a modified single peak shape for this solar cycle was suggested by a number of authors \citep[see, for example][]{usoskin2004, Vaquero2007a, Vaquero2007b, Vaquero2014} after more historical records of sunspot counts were discovered. Although, the derivation by \citet{Clette2014} indicated that the real shape of cycles in early years is not yet confirmed.

Then an almost 400-year history of sunspot activity from 1610 to the 2000s was revised by joint efforts of researchers  \citep{Svalgaard2016, Svalgaard2017a, Svalgaard2017b}. The project used on sketches of sunspots by Christoph Scheiner, Johann Kaspar Staudacher, Heinrich Schwabe, Rudolf Wolf, and Hisako Koyama \citep{Carrasco2020, Hayakawa2020}. The authors used two backbones: the Schwabe (1794 – 1883) and the Wolfer (1841 – 1944) backbone \citep{Svalgaard2016}. 
Since the two backbones have an overlap by 42 years so they were cross-calibrated with confidence. 

\citet{Svalgaard2016} (SS16 hereafter) has recounted the groups (and spots) for the present analysis and they use that recount as the base against which to normalize the counts by other observers.  In order to bridge the gap with a poor overlap between the Schwabe and Staudach backbones and by examining the data for the decades surrounding the year 1800 when the change of the lead observer happened, \citet{Svalgaard2016} concluded that the group counts reported by these observers during that interval can fall into two categories: “low count” observers and “high count” observers.  The scale factor between the low and high categories is 1.58 ± 0.15, so \citet{Svalgaard2016}  scaled the low observations to the high category by multiplying by this scale factor.

Hence, it was found \citep{Svalgaard2016} that overall, Wolf undercounted the number of groups by $\approx$ 25 $\%$,  while the counts of sunspots by these authors agree closely with Wolf's one. The authors also concluded that solar activity in 20 century was not much higher than that in the 18th century  \citep{Svalgaard2016, Svalgaard2017}.  Although, the results by \citet{Svalgaard2016} for cycle -10 do not agree with that by \citet{hoyt1998b}, while \citet{Lockwood2014} found that the shape -10 was similar to that of \citet{hoyt1998b}. 
  
 Recent data revision of potential periods of sunspot maxima and cycle durations derived from the carbon $^{14}C$ isotopes in the trees is shown the shifts in 18 century of some maxima of sunspot numbers to the early years \citep{usoskin2021}. This highlights the fact that sunspot indices in the first 15 cycles are based on unreliable data or the data with many wrong attributes not known then to the person building the sunspot index,
 
Since July 2015, the SILSO International Data Center (Sunspot Index and Long-term Solar Observations) at the Royal Observatory of Belgium maintains a new, revised series of relative sunspot numbers SSN (Version 2.0),
which was considered to be fairly reliable since 1750 \citep{Clette2014, Leussu2013, Svalgaard2016}. This solar activity data SSN V2 will be used here for a further comparison.

The main differences between SSN Version 2.0 and the previous series are as follows:
\begin{enumerate}
\item the observational series of Alfred Wolfer was taken as a basis and not the observational series of Rudolf Wolf by increasing the earlier values by approximately 1.67 times, making them comparable with modern definitions;
\item the values after 1947,  since Max Waldmeier  introduced “weights” in accordance with the size of the sunspots;
\item a noticeable trend was eliminated in the solar activity index derived from the observations of the Locarno Observatory, which was a reference observatory after 1980.
\end{enumerate}

\begin{figure}
\includegraphics[scale=0.35]{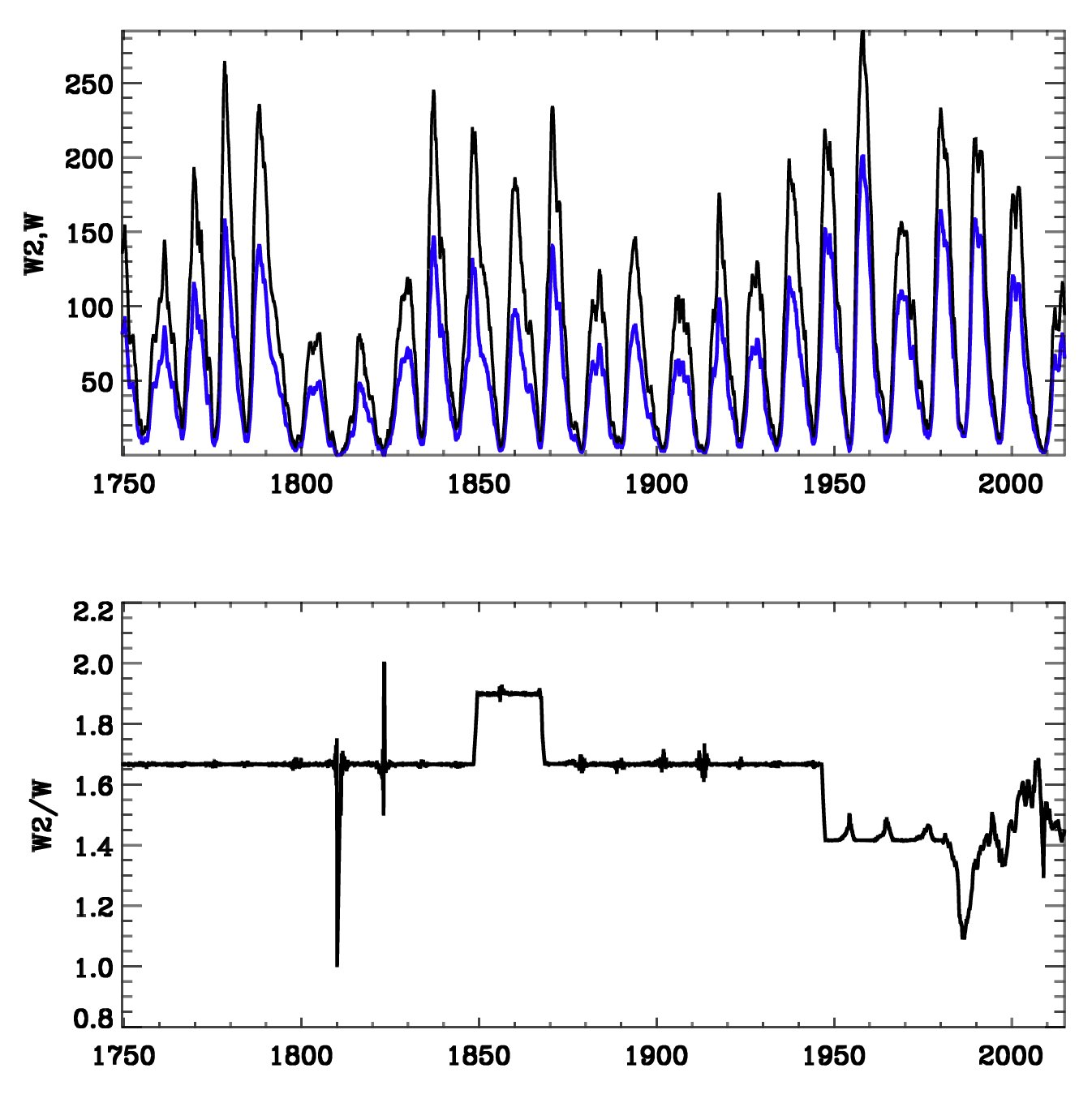}
 \caption { Top plot: Change over time of the monthly smoothed relative sunspot number  SSN (“new”, version 2.0, black line, “old”–blue line); Bottom plot:  the relationship between “new” and “old” values. A courtesy of \citet{Vasiljeva2021}. }
  \label{obs1}
\end{figure}

 The available observations of sunspots for given periods are summarised in Fig. \ref{obs1}, top plot  and  while the plots of of “new,” revised, and “old,” up to 2015, monthly smoothed sunspot numbers  as solar cycle characteristics according to SSN version 2.0 and their ratio are shown in Fig. \ref{obs1}, bottom plot \citep{Vasiljeva2021}.  The maximum number of sunspots was observed in the 19th cycle (285.0) and the minimum number of sunspots was observed in the sixth cycle (81.2). The shortest and longest cycles were respectively the second and fourth cycles with the duration of 9.0 and 13.58 years \citep{Vasiljeva2021}.

\subsection{Bayesian method restoration} \label{ssn_bayes}

As the first approach,  \citet{velasco2021} (VH21 hereafter) explored the capacity of machine learning (ML) as a powerful tool to understand the complex nature of solar magnetic activity and to predict this solar activity for the next few cycles, similar to \citet{zhar15}. VH21 used the Bayesian inference for Least-Squares Support-Vector Machines (LS-SVM) regression  \citep[see, for example,][]{suykens2002} to forecast the expected number of sunspots in the following solar cycles. 

VH21 used two models: the Non-linear AutoRegressive eXogenous (NARX) model to create models of solar cycles hind-casting and forecasting and then combine wavelet-LS-SVM algorithms for the estimation of solar cycles for a multi-channel system \citep[see section 2.3 in][]{velasco2021}. For training, validation, testing and obtaining the hyper-parameters of the model, VH21 used K-fold cross- validation algorithm  for the K values varying from 5 to 20 for the Bayesian inference model of sunspots to obtain predictions of sunspot cycles between 25 to 32.

Then in the later paper \citep[][VH22 hereafter]{velasco2022} the authors extended their model for a description of the whole set of solar activity indices defined by GSN series \citep{hoyt1998b} in cycles 1-25 and to -10 in 16-17 century. VH22 carried out a comparison of their Bayesian ML model RBg  and re-emphasized the key distinction in describing all these recent efforts to produce the best sunspot numbers for the whole series compared to the original and  focused effort based on the GSN by \citet{hoyt1998b}, on the sunspot reconstructions SSN by \citet{Clette2014}  and  \citet{Svalgaard2016} using a correction factor of 1.55 versus the correction factor of 1.208 in \citet{hoyt1998a, hoyt1998b}. 

The authors found that compared to the photographic records of the Royal Greenwich Observatory (RGO), most visual observers missed about 10 – 20$\%$ of sunspot groups. The good quality of the RGO solar images was confirmed  by the Debrecen Heliophysical Observatory \citep{Baranyi2016, Gyori2017}.  Different authors pointed out that Schwabe missed some sunspot groups, e.g  about 17$\%$ as indicated by \citet{hoyt1998b} or  about 35$\% $ as shown by \citet{Clette2014}. These missing groups led to the conclusion that the correction factor for Schwabe’s data should be 1.208 in the reconstruction by  \citet{hoyt1998b}.

A good coupling was found by VH22 between their Bayesian ML model and other models for sunspot cycles from -12 to -3. Although, they recorded a largest discrepancy with the results by \citet{Svalgaard2016} for sunspot Cycles -10, -4, and -3, noting that in the case of cycle -10 \citet{Svalgaard2016} results do not fit the reconstruction  by \citet{hoyt1998b}, while \citet{Lockwood2014} reported a very good fit for this particular cycle to the data by \citet{hoyt1998b}. According to ML reconstruction, VH22 also suggested that \citet{Svalgaard2016} may have overestimated the amplitudes of  cycles -6, -4, and -3, while the reconstructions of \citet{hoyt1998b} and \citet{Lockwood2014} seem to underestimate GSN slightly in cycles from -2 to -4.

VH22  shown that the three-peak shape (the $\Psi$ shape) of cycle -1 and unusual shapes other solar cycles (-2, -10, and -11) disappear when reconstructed with ML, and the cycles become similar to others with a single peak. VH22 suggested the unusual shapes was indirect hint that some sunspot numbers are missing in these cycles. 
Also VH22 shown that the original Wolf’s reconstruction (WSN) had a good agreement with their Bayesian model  for the period from 1800 to 1849, while this agreement worsened in 1750 -1800, when the data were sparse. 

Also VH22 reckon that the amplitude of Cycle 11 is not resolved \citep{Senthamizh2015} because of the WSN calibration before 1849. A detailed analysis by \citet{Leussu2013}, which found the GSN series of \citet{hoyt1998b} to be more consistent and homogeneous with Schwabe’s data.  VH22 confirms that the GSN reconstruction  \citep{hoyt1998b} agrees with the original, unadjusted WSN record. Although VH22 still  questioned the validity of maximum magnitude  in cycle 5 and its correct duration. 

\subsection{Problems with the manual and  Bayesian restorations} \label{probs}
\subsubsection{Estimated uncertainties of manual sunspot detection  and indexation} \label{discr}
The  known problems with the establishing  sunspots times and locations as well as deriving the solar activity index in early years:
\begin{enumerate}
\item Unexperienced drawings by some observers in early years.
\item Poor observations by some observers leafing to the difficulties to build a backbone.
\item It is unknown if the number of sunspots in groups changes over time.
\item Accounting for years with less than 20 observation days per year that brings large errors.
\item  The impossibility of determining the weighting factor for the periods with a single observer.
\item There were different calendars  (Julian versus Gregorian) used until mid or late 18 century in different countries where the dates placed on drawings did not indicate a real calendar iused.
\item In a few key countries there were different starts of a New Year (April in Britain, September in Russia etc) used meaning that unless recorded, the year on the sunspot drawings also would not correspondent to the accepted calendar year.
\end{enumerate}

 While the items 1-6 were actively discussed numerous times in the most papers cited above exploring various solar activity indices, the items 7 and 8 were avoided from this attention while they can contribute with the major errors in defining solar cycles maxima and durations.

Recent data revision shown the shifts of maxima of sunspot numbers to the early years (about 18 century) towards the years originally assigned to minima \citep{usoskin2021}. This highlights the fact that the sunspot indices in the first 15 cycles of 17-18 centuries could be build using both simply malign data and the data with many wrong attributed times  not known then to the person building the sunspot index.

\subsection{Discrepancies after fitting the sunspot indices with Bayesian approach} \label{ssn_discr}
There are two reconstructions VH21 and VH22 of the solar cycles represented by sunspots produced by the same group of authors \citep{velasco2021, velasco2022}, which reveal slightly different temporal patterns.

\begin{figure}
\includegraphics[scale=0.24]{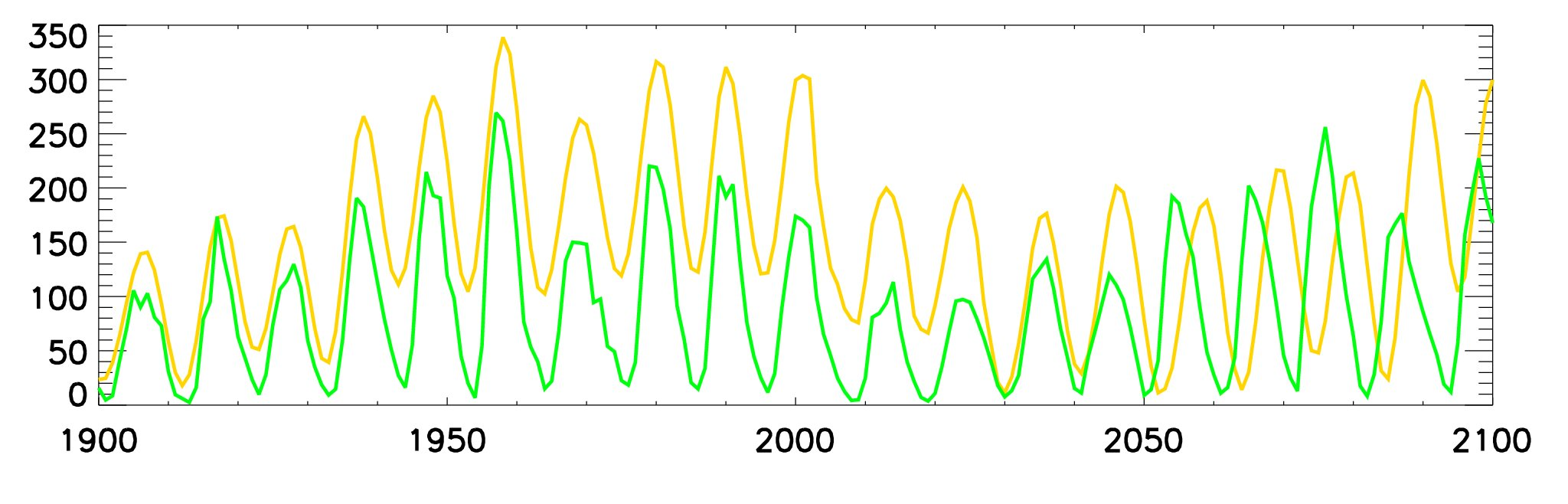}
 \caption{A comparison of the PMOD reconstruction with Bayesian approach (orange line) \citep{velasco2021} with the sunspot reconstruction (green line) \citep{velasco2022}. }
\label{pmod-bayes}
\end{figure}

Their fitting really depends on, which part of the sunspot data the Bayesian approach was focused on, as shown in Fig. \ref{pmod-bayes}. In particular, if the the Bayesian model was fitted  to the cycles from 1850 towards the modern times (VH21) the Bayesian model shows to the modern grand solar minimum followed by the next grand solar cycle, as it was predicted by \citet{zhar15}. If the Bayesian model was fitted by the whole series of sunspot cycles VH22, then the solar activity curves in cycles are shifted to earlier times as shown in Fig. \ref{pmod-bayes} by comparing the green (VH22) and yellow curves (VH21). Although, the green curve VH22 shows good  fitting of Maunder minimum and most of the cycles while struggling to fit some very early ones when there were little observations and the latest solar cycles in 20th century.

Hence, in addition to the general problems with building the averaged sunspot numbers listed in the previous section,  there are remaining problems affecting the sunspot index if fitted by the Bayesian methods.
Bayesian fitting centred either to 18-19 century data or towards 20-21 century data produced the arrays, which do not exactly coincide in the intermediate points. This correlates with the recent data revision of sunspot maxima and cycle durations with carbon 14 isotope data  showing  in the years in 18 century  some shifts of the maxima dates  to the dates where originally were minima.
These discrepancies could be caused by problems with the sunspot identification built from the early data as follows.

\begin{enumerate}
\item Poor or none observations by some observers in the whole 18th century.
\item Difficulties to build a backbone for the beginning of 19 century still affect fitting the solar activity curve.
\item Different calendars in different countries (Julian-Gregorian until the mid 18th century), different starts of a new year in some countries (April in Britain, September in Russia etc) can significantly affect timing of sunspot numbers and lead to wrong cycle durations and maxima.
\end{enumerate}

The problems with early cycles in 18 century  are likely caused by the absence or scarceness of the observations in 18 centuries (see Fig.\ref{obs}), which cannot be repaired unless more historical data are found. Furthermore, in the 18 century many countries were moving from Julian calendar to Gregorian one  in very different times (see for example, the table here 
$https://en.wikipedia.org/wiki/List_of_adoption_dates_of_the_Gregorian_calendar_by_country$. These calendar moves happened in different years in different countries and not always the  calendar used is indicated in the drawings used for the sunspot index definition. Therefore, reliabity of the sunspot data  in these years is rather questionable.

To complicate the matters the different countries had different starts of their years. In Britain the new year was starting in April  until 1852, after which they accepted the start of New year in January 1853. This meant that, for example, the drawings made in January-March 1750 by the old calendar are actually belonged to the next year 1751 and so on. The similar discrepancies could occur with the drawing carried in the country where the year start was September or the other months.  These objective problems led to some very strange solar cycles derived for the 18th century, when the cycle lengths were  either 7 or 8 years while other cycles lasted for 15 years.

These discrepancies raise more questions to the quality of solar activity index defined by the sunspot data  used currently by many restorations as the sunspot cycles in 18-19 centuries. Nonetherless, a use of this familiar sunspot indices, either Wolf numbers  WSN or sunspot numbers SSN, are well accepted by the solar-terrestrial community, despite any discrepancies and difficulties they experience with the replication of these solar activity (SA) indices. We fully appreciate tremendous efforts of the solar community to this topic, which helped researchers to understand the nature  and need of solar activity indices and to improve their accuracy in the modern days as far as it is feasible with a sunspot proxy. 

However, for the sake of diversity, it is worth to look at some other possibilities to define solar activity, like Bayesian models shown here (VH21, VH22) or to use  Principle Components Analysis  index recently suggested  \citep{zhar15, Zharkova2022} for a comparison  with the accepted index of solar activity with the averaged sunspot numbers.

\begin{figure}
\includegraphics[scale=0.82]{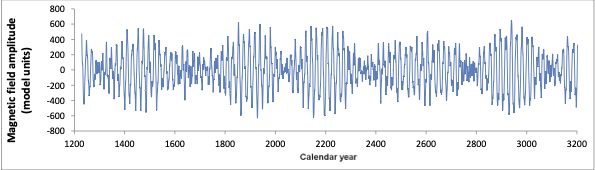}
\includegraphics[scale=0.55]{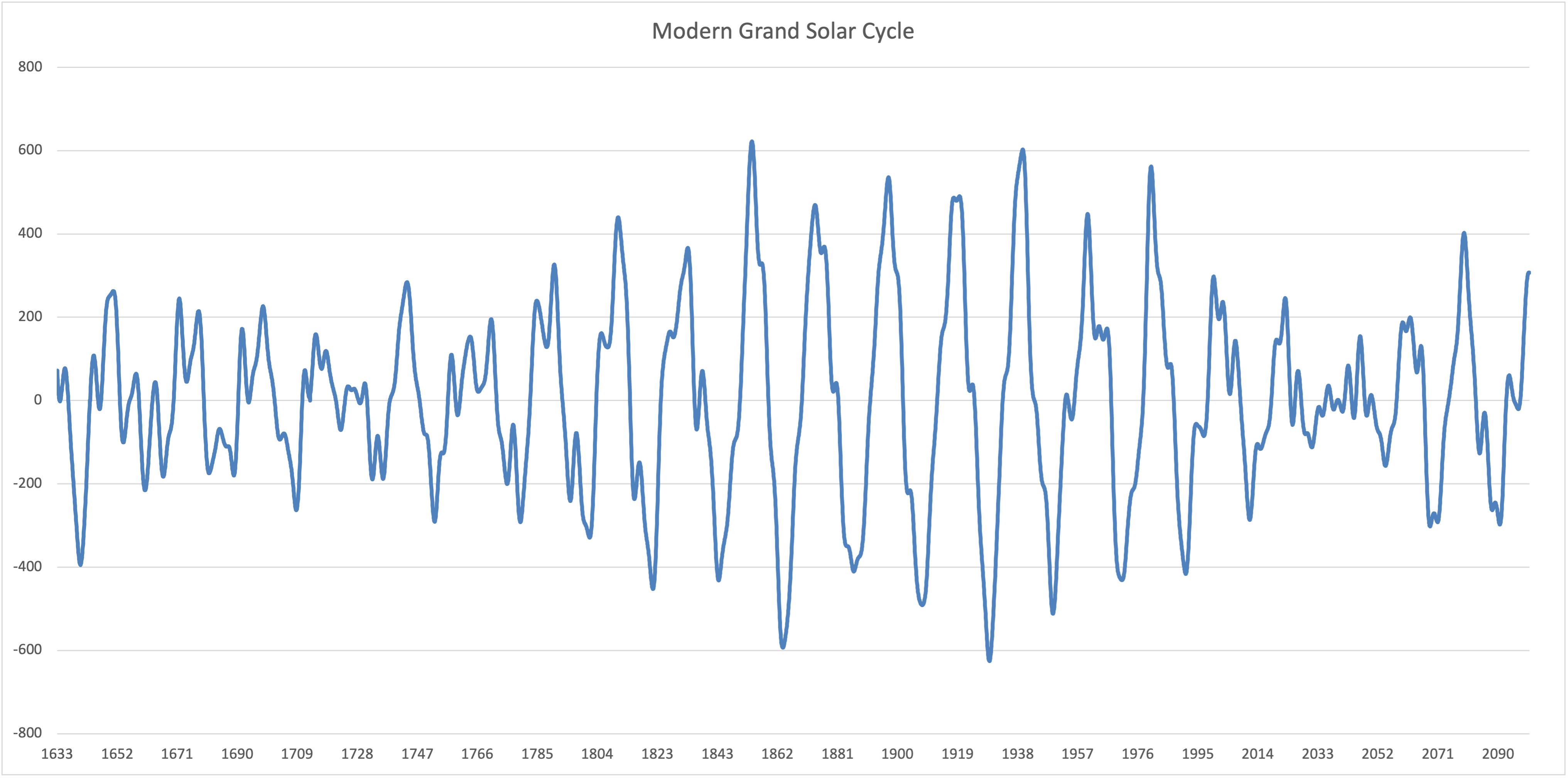}
 \caption{Top plot: The summary curve of two PCs calculated for 2000 years \citep{zhar15}. Bottom plot: the summary curve in the current grand solar cycle of 370 years.  }
  \label{sum}
\end{figure}

\begin{figure}
\includegraphics[scale=0.5]{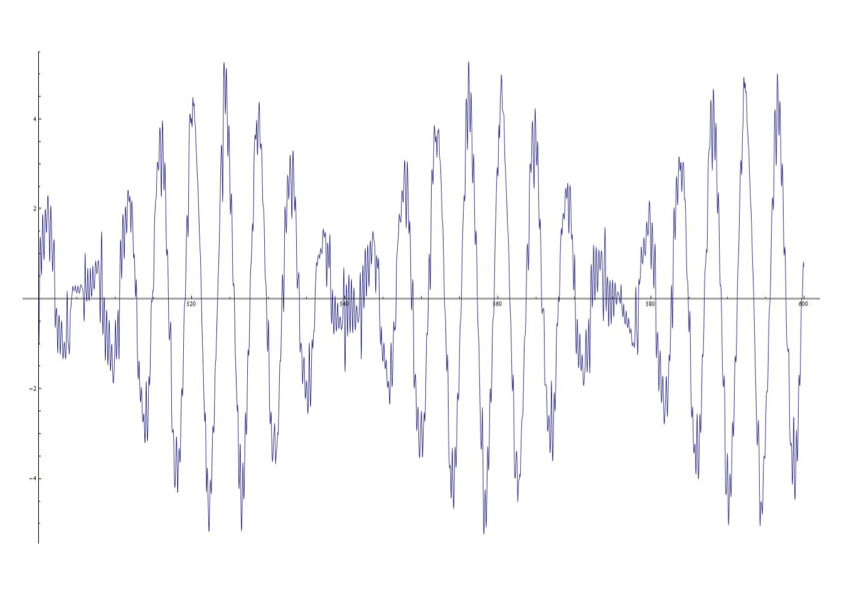}
\includegraphics[scale=0.5]{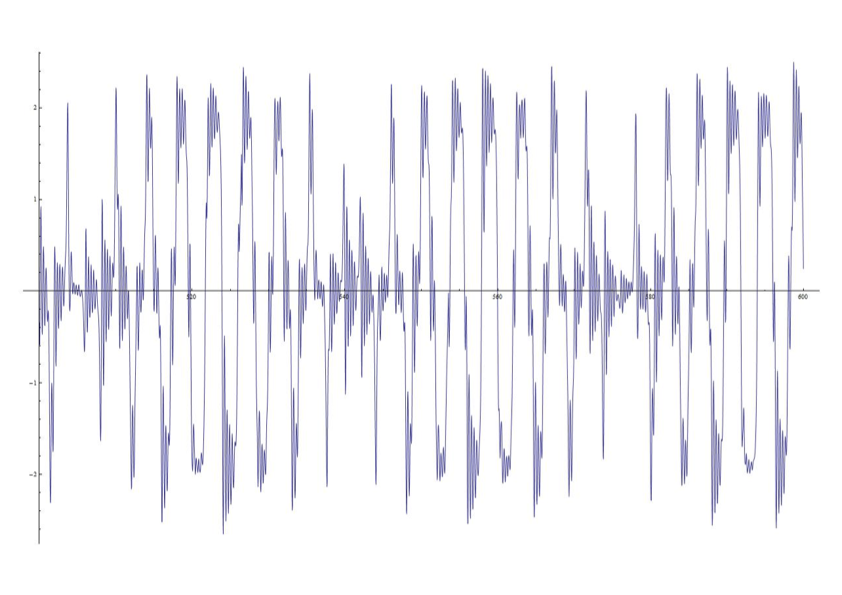}
 \caption{The simulated poloidal (top plot) and toroidal (bottom plot) magnetic fields generated by the solar dynamo for 1000 years covering the modern grand solar cycle in the centre. }
 \label{pol_tor}
 \end{figure}
 
\begin{figure}
\includegraphics[scale=0.77]{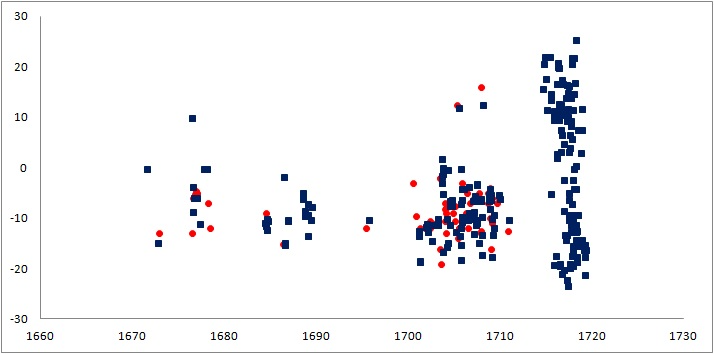} \\
\includegraphics[scale=0.58]{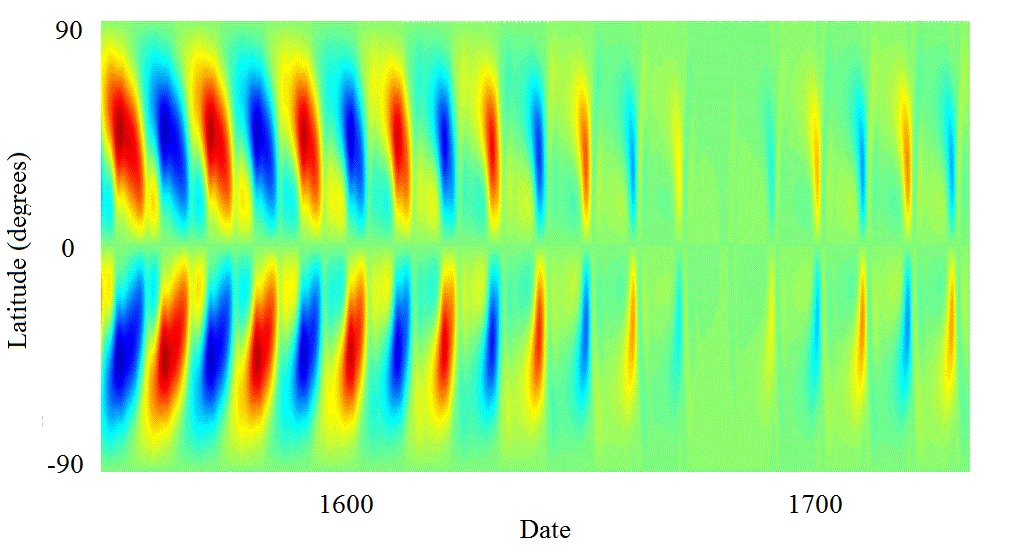} \\
\includegraphics[scale=0.58]{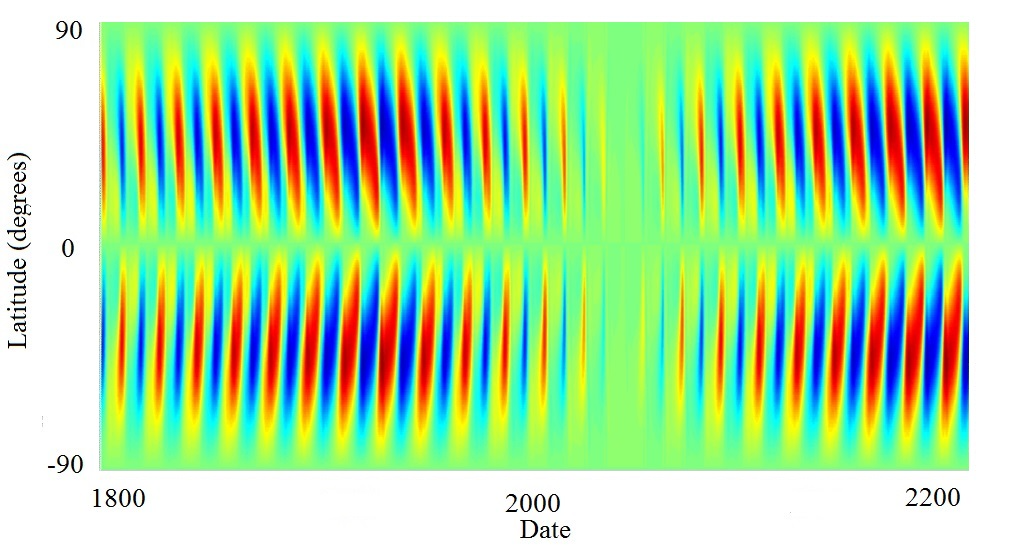}
 \caption{The butterfly diagrams derived from the observations of Maunder minimum (top plot) \citep[blue i red dots taken from][respectively]{Eddy1976, Ribes1993}. The butterfly diagram simulated using the two layers dynamo model \citep{zhar15} for the current grand solar cycle including the Maunder minimum (middle plot). The butterfly diagram simulated for the modern GSM (2020-2053) using the same dynamo model \citep{zhar15} (bottom plot). }
  \label{butterfly_mm}
\end{figure}
 
\section{Solar activity index from the eigen vectors of solar magnetic field} \label{ss_pca}
\subsection{Pair of eigen vectors, or principal components} \label{ev_sum}
\subsubsection{PCs and their summary curve} \label{ev_proxy}
 The dynamo mechanism, which governs solar activity, operate with poloidal and toroidal magnetic fields \citep{Parker55}, with the first one being  the solar background magnetic field (SBMF), and the second one being the magnetic field of magnetic loops in active regions, which are embedded into the solar surface, whose roots are seen as sunspots.  The interaction between these two magnetic fields defines the variations of solar activity seen through the appearance or disappearance of sunspots and active regions.
 
  However, because the SBMF is shown to be in anti-phase with the leading polarity of magnetic field in sunspots \citep{Stix76, zharkov08} defining the locations and timing of sunspot appearances on the solar surface  and their migration towards the solar  equator or poles \citep{zharkov08}, one can expect that the summary curve of the SBMF should reveal defined links with the averaged sunspot numbers.  
    
 \citet{shepherd14, zhar15} investigated these two principal components (PCs), or eigen vectors, of the solar background magnetic field  (SBMF) by applying the principal component analysis (PCA) to  the Wilcox Solar Observatory low resolution full disk synoptic magnetic maps for cycles 21-23. The authors identified numerous eigen values and eigen vectors of  own magnetic waves of the Sun seen on the solar surface, which came in pairs. The four significant pairs covering the majority ($>95\%$ of the data by variance \citep{zharkova12, Zharkova2022}. 
 
The first pair, or two principal components, reflect the primary waves of solar magnetic dynamo produced by the dipole magnetic sources \citep{zhar15}. These two waves are found traveling slightly off-phase  from one hemisphere to another  and their interaction define the solar activity in each hemisphere and as a whole \citep{zharkova12}.    \citet{shepherd14, zhar15} used the symbolic regression analysis  \citep{Schmidt09} of these two magnetic waves and obtained the analytical expressions for the magnetic (dynamo) waves incorporated into  the ensemble of waves present in the solar background magnetic field  attributed to the poloidal field of the Sun \citep{popova13}. 

These mathematical equations were used to make predictions in time by thousand years both forward and backward, from the current epoch and to use them for a comparison with the magnetic waves supposedly produced by the solar dynamo acting in two layers  with slightly different meridional circulation velocities \citep{zhar15}.  

In order to bring the  detected trends in the SBMF closer to the currently-used index of solar activity, the averaged sunspot numbers, we calculated the summary component of the two PCs. The modulus summary curve was found to correlate closely with averaged sunspot numbers \citep{shepherd14, zhar15}. \citet{zhar15} suggested to use  the summary curve of these two PCs as a new proxy of solar activity, instead of, or in addition to, the averaged sunspot numbers. This suggestion was confirmed by the previous research with PCA  of the solar magnetic synoptic maps of Kitt Peak observatory obtained with much higher resolution \citep{Zharkova2022}, which PCs, in fact, reflect the magnetic fields of active regions, e.g. toroidal magnetic field, well known to have the  11 year periodicity classified via averaged sunspot numbers. While the lower resolution  SBMF from WSO used in our previous research \citep{zharkova12, zhar15, Zharkova2022} PCA detects the two PCs, or magnetic waves, of the poloidal magnetic field associated with dipole magnetic sources \citep{zhar15} and another three significant pairs associated with quadruple, sextuple and octuple magnetic sources \citep{Zharkova2022}.

Using the derived formulae, the summary curve was calculated backward  to 1200  and forward to 3200 \citep{zhar15} as shown in Fig. \ref{sum}, top plot  revealing very distinct variations of the cycle amplitudes in every 350-400 years, or grand solar cycles. These grand solar cycles are separated by grand solar minima (GSMs),  when the amplitudes of 11 year cycles become very small, similar to those reported in Maunder,  Wolf and Oort  and other grand solar minima \citep{Zharkova2018b, Zharkova2018a}. The current grand solar cycle (GSC) shown in Fig. \ref{sum}, bottom plot started during Maunder Minimum and will continue until  the modern GSM (2020-2053) as  pointed by \citet{zhar15}.

The timings of the grand solar minima are defined by the interference of two magnetic dynamo waves generated in different layers with close but not equal frequencies  defined by the different velocities of meridional circulation (so called beating effect) \citep{zhar15}.  The calculation of the summary curve  forward in time until 3200 has  shown the further three grand solar cycles separated by three GSMs with the first GSM  to occur right now during the cycles 25-27, or in 2020-2053 \citep{zhar15, zhar2020}. 

\subsubsection{Modulus summary curve and the sunspot index} \label{ev_ssn}

 The temporal features of a modulus of the summary curve of these two PCs shown a remarkable resemblance to the sunspot index of solar activity for cycles 21-23 \citep{shepherd14, zhar15}, or recently cycles 21-24 \citep{Zharkova2022}.    Embracing the similarity between modulus summary curve and averaged sunspot numbers, the modulus summary curve can be normalised for each cycle by the averaged sunspot numbers indicated by the left Y-axis. The modulus curve, in general, follows the averaged sunspot numbers for all the the cycles revealing a significant reduction of solar activity from cycle 21 (maximum about 300 sunspots), through cycle 22 (230), 23 (165)  to cycle 24 (108) \citep{Zharkova2022} that fits reasonably to the maximum numbers reported for cycles 21-24 \citep{SILSO}: 21 - 233, 22 - 213, 23 - 180, 24 - 116.
 
Although there was a remarkable resemblance between these two curves, given the fact that they represent different magnetic components of solar dynamo waves: poloidal for the modulus summary curve and toroidal for averaged sunspot numbers.   This similarity despite the summary curve reflects the poloidal magnetic field while sunspots - toroidal one allowed  them to suggest this summary curve of PCs, or eigen vectors of SBMF, as a new solar activity proxy. The advantage of using  the solar index from the summary curve instead of the averaged sunspot numbers is the ability to do long-term predictions and a presence of the extra-parameter, a leading polarity of the background magnetic field of the Sun.  
  
  Hence, from the one hand, the modulus summary curve is found to be a good proxy of the traditional solar activity index  contained in the averaged sunspot numbers. This suggestion was also supported by the recent research of the same SBMF data of WSO \citep{Kitiashvili2020, Obridko2021}. At the same time, the summary curve  proposed by \citet{shepherd14, zhar15}  as a sum of the largest eigen vectors of SBMF, which were given mathematical description via a series of cosine functions, is shown to represent a real physical process - poloidal field dynamo waves generated  from dipole magnetic sources by the solar dynamo in two layers of solar interior \citep{zhar15}. 
 
  Therefore, the modulus summary curve (MSC) proves that the eigen vectors of SBMF can be considered as a very good proxy of the traditional solar activity understandable by many observers. It should be used as complementary solar activity index in addition to the existing one of the averaged sunspot numbers. Plus, this new index adds the additional parameter to this proxy - a dominant polarity of the solar background magnetic field for each cycle, which has the  polarity opposite to the leading polarity of sunspots \citep{Stix76, zharkov08}.

 Based on the similarity of the modulus summary and sunspot curves, one can conclude that the solar activity in cycles 21-24 is systematically decreasing with cycle number \citep{zhar15, Zharkova2022} because of the shift in phase of the two magnetic waves so that their phase difference  is increasing in time, approaching an anti-phase when there is a lack of any interaction between these two dynamo waves. This wave separation into the opposite phases will definitely lead  to the absence of active regions, or magnetic flux tubes, whose roots  appear on the solar surface as sunspots. This, in turn, can lead to am absence of noticeable activity on the solar surface, especially, in the descending and minimal phases of cycles 25-27 \citep{zhar15} that can resemble the similar features recorded during Maunder Minimum  \citep{Eddy1976}.

 \subsubsection{Links  to the waves of solar dynamo} \label{sec:tor_pol}
   
   The previous simulations of poloidal and toroidal magnetic fields of the Sun from Parkers two layers dynamo model with meridional circulation \citep{zhar15} produced by dipole magnetic sources \citep[see Fig. 6, top plot  in][]{zhar15} revealed their very close correspondence  of the simulated magnetic waves for poloidal  magnetic field to the derived principal components and their summary curve for 2000 years shown in Fig. 3 of \citet{zhar15}.  
   
   For the current grand solar cycle from MM until the modern GSM (see Fig. \ref{sum}, bottom plot) the simulated poloidal and toroidal magnetic fields using the dynamo models described in paper by \citet{zhar15} are shown in Fig. \ref{pol_tor}, top and bottom plots, respectively.
 These plots visibly demonstrate that the appearance of poloidal and toroidal magnetic fields are very different. The amplitudes of poloidal magnetic field in the current grand solar cycle  (GSC) (1685-2043)  varies very significantly with time of the GSC cycle while the amplitudes  of toroidal field  are much less changeable during the GSC  and they only decrease towards grand solar minima. Therefore, any proxies of solar activity associated with these two magnetic field are designed to be somehow different. Note, in this simulations we did not include any centennial oscillations of magnetic field (Gleisberg cycle) that could affect the amplitudes of the toroidal magnetic field. 
 
 In addition to the temporal variations of  poloidal magnetic field  we can also simulate the butterfly diagrams measured for the current grand solar cycle including the Maunder minimum (MM)  shown in Fig. \ref{butterfly_mm}, top plot revealing a very limited sunspot formation during the MM years.
 This butterfly diagram compared with the butterfly diagrams simulated for the current grand solar cycle with the same dynamo model as described in \citep{zhar15}. including the butterfly diagram about Maunder minimum (MM) (see Fig. \ref{butterfly_mm}, middle plot) and the butterfly diagram for the modern GSM (see Fig. \ref{butterfly_mm}, bottom plot) .
 
 It can be noted  that the  butterfly diagram simulated for the current grand solar minimum with the same dynamo model as described in \citep{zhar15} closely repeats the pattern of the observed butterfly diagram during the long grand solar minimum, MM, lasting for 6 solar cycles. 
 Furthermore, the simulated butterfly diagram for MM resembles the patterns of magnetic field (of sunspot) occurrences at different latitudes in the years of nearly absent solar activity (1645-1700) and then towards the end of GSM  and after it in the years  1700, 1710 and 1720. 
 
 The model of the butterfly diagram for the modern GSM reveals the absence of sunspots in the butterfly diagram only during cycle 26, while there are some remaining magnetic activity is observed during cycles 25 and 27. This indicates that the modern GSM (2020-2053) will be shorter and slightly more active than the MM GSM.
 
 \begin{figure}
\includegraphics[scale=0.3]{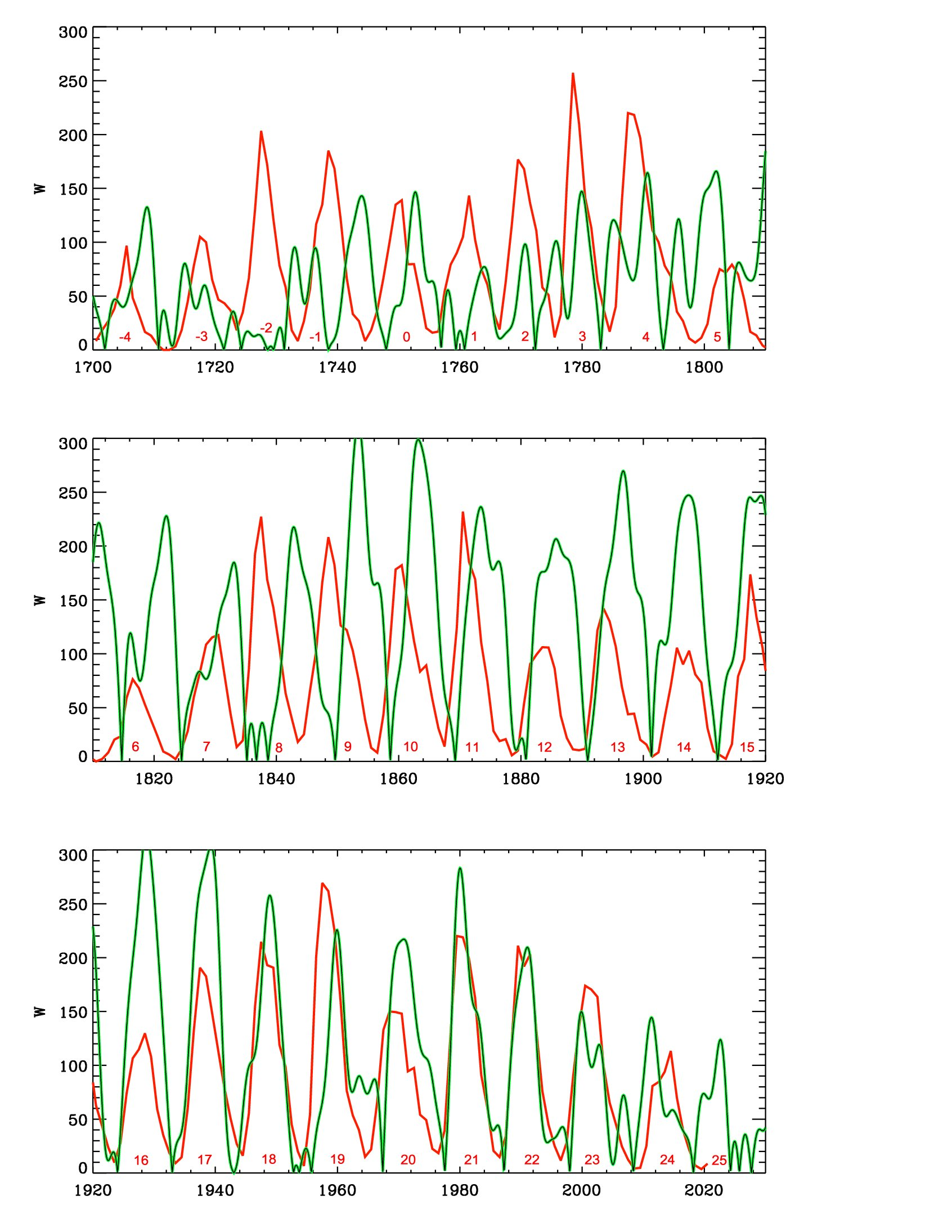}
 \caption{Comparison of the modulus summary curve (green) \citep{zhar15} with the averaged sunspot numbers (red) \citep{SILSO}. }
  \label{msc}
\end{figure}

 \begin{figure}
\includegraphics[scale=0.59]{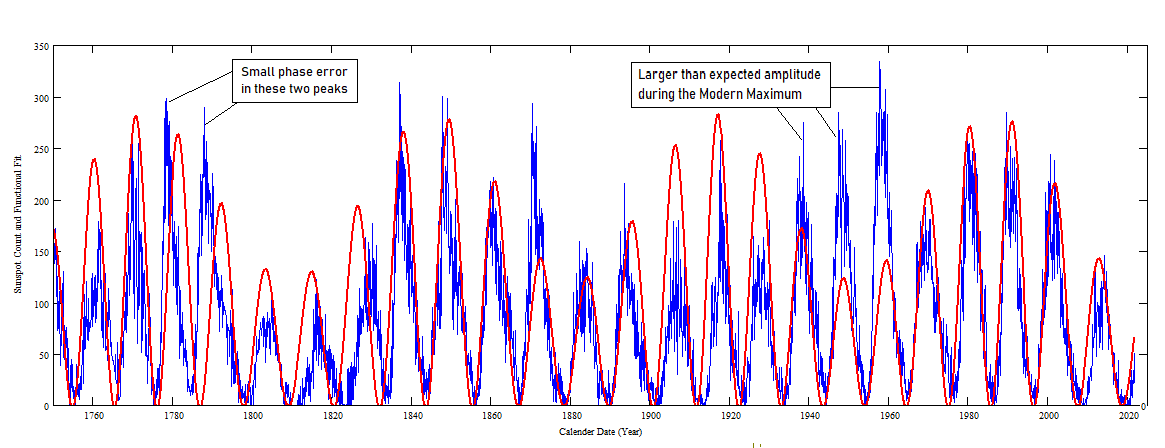}
\includegraphics[scale=0.53]{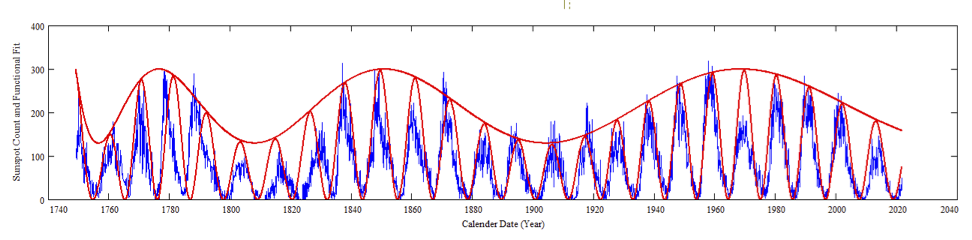}
  \caption{Fitting with regression approach the modulus summary curve \citep{shepherd14, zhar15} with the amplitudes of the averaged sunspot numbers  (top plot) and with the centennial  cycles of 89 and 112 years (bottom plot). See the text for details. Sunspot numbers \citep{SILSO} are plotted by navy lines, MSC  \citep{zhar15} by red lines. The envelope curve in the bottom plot represent overall variations of the sunspot cycle amplitudes. }
  \label{msc_fit}
\end{figure}

 \subsection{Comparing  the modulus summary curve (MSC) and averaged sunspot numbers} \label{ss_msc}
 In this section we present  a comparison of the modulus summary curve (green line) \citep{zhar15} with the averaged sunspot numbers (red line), from \url{https://wwwbis.sidc.be/silso/} \citep{SILSO} as shown in Fig.\ref{msc}.
It turns out that in the past 370 years  of the current grand solar cycle there is a rather reasonable agreement between the MSC and SSN as in the duration and maximum magnitudes for cycles 12-24 as shown in the bottom and middle rows of the plots in Fig.\ref{msc}.  Although, the maximum amplitudes for cycles of MSC curve become slightly exceeding the SSN maximum magnitudes from cycle 12 to cycle 17 that is similar to the reconstructions reported by \citet{solanki04, Chatzistergos2017}.

Then in the MSC curve cycles 8-11  (Fig.\ref{msc}, middle plot) are shifted forward from the SSN curve, slightly for cycle 11, more for cycle 10 and by a half of a cycle length for cycles 9 and 8, so that the maxima in cycle 8 and 9 in MSC occur during the minima in SSN. These are followed by correct durations for cycles 7 and 6, appearance of an additional cycle between cycles 5 (which in SSN was of 15 years duration) and 6, shifted duration for cycle 5, close durations between MSC and SSN  for cycles 1 and 4 with triple maxima in some of them. 
Then there are MSC cycles slightly shifted forward  compared to SSN in cycles -2 to 0 and close resemblance of MSC with SSN for cycles -3 to -4 followed by the Maunder minimum. 

More complicated relations with the MSC, or eigen vectors, defined from the poloidal magnetic field appear for three cycles 8-10, cycles -2 to 1 and  for cycles 4-6 and possibly 7. Cycles -2 and -3 and, possibly -1 in MSC have shown three-peak shapes, similar to those reported by \citet{Clette2014} from the reconstruction by \citet{hoyt1998b}. Definitely, cycles  3-5 in the MSC curve have two maxima shapes that followed then by a reduced second maximum in cycles 6 and 7 returning  to single maximum shapes in further cycles from cycle 8 onward.

Cycles 7 and 8 in MSC have the duration similar to the sunspot cycles but shifted forward. In cycles 8-10 minima and maxima in sunspot index were over-lapped the with maxima and minima of the MSC cycles. Then from cycle 11 the MSC shifts ahead the sunspot cycle by a year or by year and half and then follow the cycle durations of sunspot cycles being though higher in amplitudes. 

 The major discrepancies occur in the periods of  1730-1780 in 18 century and in the period of 1830-1870 in 19 century. In the period of 1725-1735 the eigen vector MSC has minima while the sunspot index SSN  shows maxima. Keeping in mind the  lack of observations in the period of 1720-1760 in the 18th century and some unreliable observations in the period of 1830-1860 in 19th century discussed in Fig. \ref{obs} in section \ref{ss_hist}, we will try to understand  and explain the differences between the MSC index and the sunspot index for these two periods where the indices are strongly different.

The most puzzling is the distribution of sunspot cycles in the 18th century that is denoted by all the investigators listed above. This is the century when some cycles had durations of 7 or 8 years while others lasted for 15 years. Contrary to this description, the MSC curve reveals in this period pretty regular cycles with double maxima ( cycles 1-4), again $\Psi$ type distribution for the shapes of cycles 0 and 1 and for cycles -1 and -2 s before they come to Maunder minimum shown in Fig. \ref{sum}  (top plot). The MSC cycles reveal smaller maximal magnitudes  in cycles -3 to 0 and in cycle 1-4 than the amplitudes of sunspot index. Again cycles -2 to 0 have a reversed maxima and minima between sunspot and MSC curves.
 
 Similar to the approach reported by SH22, let us make another attempt to fit  the parameters of the modulus summary curve to the observed averaged sunspot numbers SSN \citep{SILSO}. The fitting was carried out using Hamiltonian regression approach applying the fitting  by  a few parameters used to quantify the Principal Components \citep{Schmidt09} for a few periodic functions (cosines) and period phases by time and cycle length defining cycle periods and by amplitudes defining the maximal amplitudes, or cycle power.
  Fitting the modulus summary curve MSC  to the existing averaged sunspot numbers SSN is shown in Fig. \ref{msc_fit}.
  
It was found that this fitting is limited depending on the part, which we want to fit in the sunspot index. This problem is similar to the problems found in Bayesian approach  if one compares the results reported by VH21 and VH22  shown in Fig. \ref{pmod-bayes}. Namely, if we fit the MSC curve to the averaged sunspot numbers SSN for all cycles of the current GSC, the MSC curve fits cycle duration and maxima times but it fails, as result,  to hit the correct dates either for Maunder minimum occurred before the modern grand solar cycle (GSC)  start, or for the modern grand solar minimum (GSM) on the other side of the GSC. While, if we leave the eigen vectors of the SBMF as they are derived from the full disk magnetic synoptic maps, the timings of the GSMs are correct, while the amplitudes or positions of the individual cycles do not fit to the amplitudes and some positions of sunspot cycles in early years before 1850. 

These differences in the early solar cycles in the 18th century can be explained by a few reasons. One reason is a difference between physical entities represented by modulus summary curve, or eigen vectors of poloidal magnetic field of the Sun and loose representation of toroidal field by the averaged sunspot number.poor observational data for sunspots reported by all the papers. These difference can reflect the real differences between the cycles in toroidal (sunspots) and poloidal (SBMF) magnetic fields as discussed in section \ref{sec:tor_pol}. The discrepancies between the maximum amplitudes and some shifts of maximum times can be understood in terms that the MSC is produced by poloidal magnetic field while the averaged sunspot numbers are loosely associated  with toroidal magnetic field, though this link is rather distant via the number of sunspots in groups.
 
 Another reason for the discrepancies and shifts of MSC cycles by 5-6 years can be a poor coverage of the observations during the early years as discussed in section \ref{discr} and shown in Fig. \ref{obs}. The contribution to these problems can be also mixed not only with the physical absence of observations but also with a big misunderstanding in dating accurately the sunspot drawings because of the calendar change  from Julian to Gregorian at different times in different countries, or with the varying starts of New Years in different countries that could also make a complete mess  shitting the dates assigned for the sunspots by more then a year.

There is also a possibility of uncertainty of the detected eigen vectors of SBMF used for PCA \citep{zhar15}. However, the most recent investigation of eigen vectors of SBMF from cycles 21-24 \citep{Zharkova2022} has revealed that the addition of the magnetic field data  from the extra solar cycle 24 did not change the eigen values and eigen vectors, at least, for the first four pair, or eight eigenvectors. This is a reassuring finding, which strengthens the case for using  the eigen vectors of poloidal field as a good proxy of solar activity, in addition to the existing solar indices defined by averaged sunspot numbers or by Wolf numbers.

 \subsection{Comparing the sunspot indices and restorations  in the current grand solar cycle}\label{ss_early}
 \begin{figure}
\includegraphics[scale=0.17]{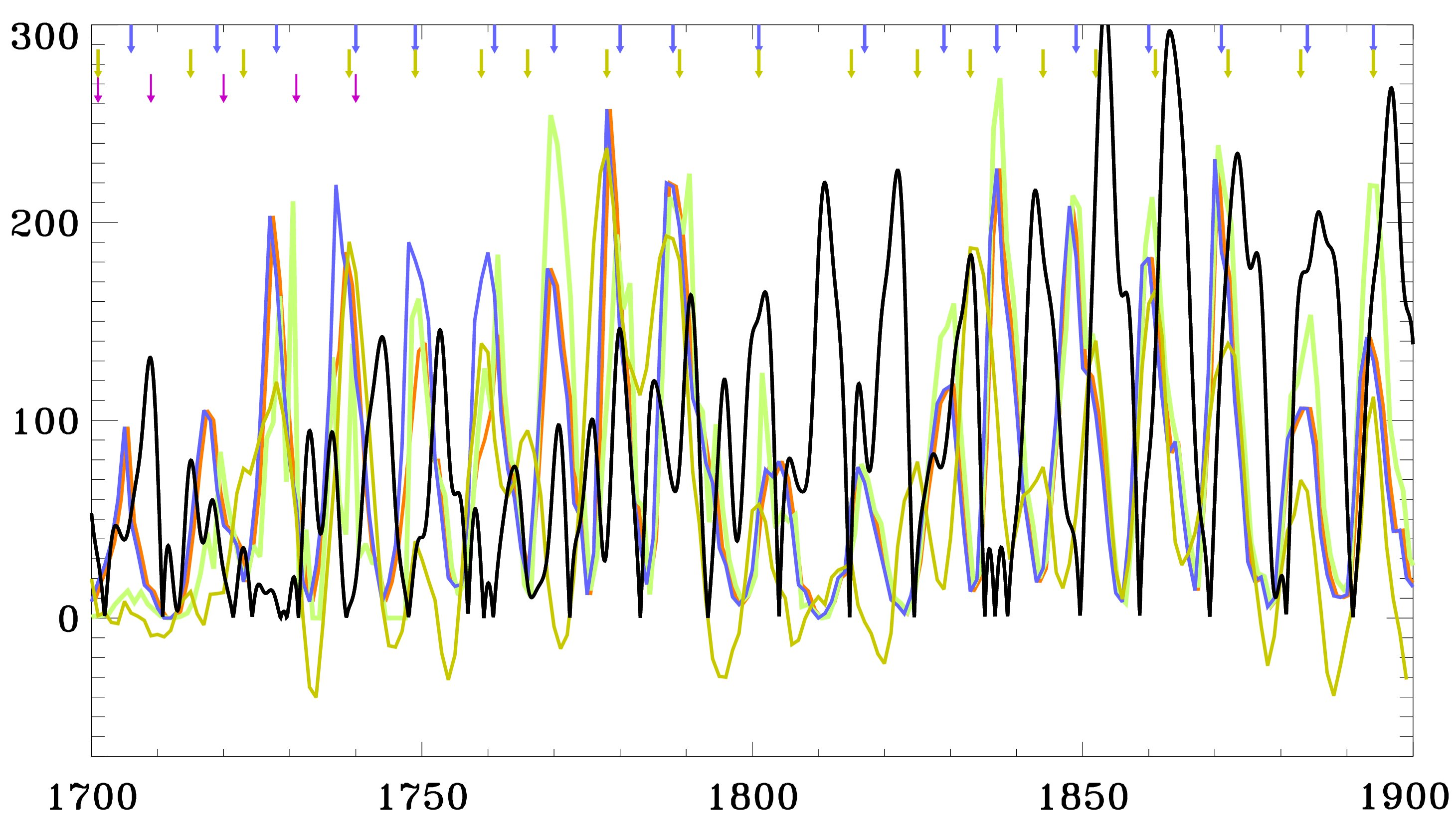}
  \caption{The comparison of sunspot restoration  by different authors including from the missed observations \citep{usoskin2021}. The arrows on the top indicate the maxima derived by the relevant restoration. The lines represent the following: averaged sunspot number SSN \citep{SILSO} (red),  summary curve \citep{zhar15} (black line), HS98 GSN \citep{hoyt1998b} (green), Bayesian restoration \citep{velasco2022} (blue), recent restoration \citep{usoskin2021} (mustard).  The arrows defining the sunspot maxima by different authors are described in the text.}
  \label{ss_max}
\end{figure}

For better understanding the solar indices let us compare their restorations done by various authors. The plots for a few reconstructions of solar activity, or sunspot index, are shown in Fig. \ref{ss_max}, demonstrating the locations of maximum solar activity according to different authors. A comparison of the solar activity cycles and newly found maxima of solar activity derived from the updated data of sunspots  by different authors are shown in Fig. \ref{ss_max}. The arrows indicate the maxima suggested by different authors: red arrows by \citet{Beer1998},  grey arrows by \citet{Maunder1922}, green arrows by \citet{waldmeier61}, orange arrows by \citet{Schove1983}, blue arrows by \citet{velasco2022} and mustard arrows by \citet{usoskin2021}.

It can be seen a very interesting development reported in the recent data revision with the new sunspot activity restoration derived from the abundances of the isotope $^{14}C$ in the trees \citep{usoskin2021}.  The authors shown y the shifts of the maximal of sunspot numbers in the 18th century to the years  where originally there were minima of sunspots. It is clear that the discrepancy between the different data on the maxima of solar activity sometimes reaches 5-6 years \citep[see also discussion in ][]{velasco2022}. These shifts of the SSN maxima make them close to the maxima of modulus summary curve (MSC) reported from PCA  \citep{zhar15} for this period for cycles from -2 to ,  and for cycles  8-11as shown in Fig. \ref{msc}, top and middle plots.

Keeping in mind that the summary curve represents the principal eigen vectors of the solar background magnetic field while the SSN or GSN series represent some sort of toroidal field but not exactly as they include the numbers of sunspots but not their magnetic fields. This difference between the properties of the series can explain the difference in the amplitudes of cycles -2 to 4 in the interval closer to Maunder minimum where the amplitudes of poloidal field, or MSC, is much lower than that of SSNs and for cycles 9-17 where the poloidal magnetic field, or MSC, magnitudes were much higher.  In addition to any observational restrictions discussed in section \ref{probs},  these differences between MSC and SSN curves could  also indicate the real difference between poloidal and toroidal magnetic fields in these periods linked to some specific interaction of two magnetic fields in the solar dynamo model as discussed in section \ref{sec:tor_pol}. 

 \section{Discussion and conclusions} \label{conclusion}
For many decades  researchers used to evaluate solar activity through variations of sunspot numbers. They tried to link these variations to the action of solar dynamo, e.g. processes of generation of magnetic loops (toroidal field) in the solar interior, their transport to the solar surface and consequent disintegration of these loop magnetic fields into the background magnetic field (poloidal field) of the Sun  (SBMF) \citep{Parker55}.

Tremendous work has been done by numerous researchers to setup the current solar activity index - the averaged sunspot numbers \citep{SILSO}. \citet{Schwabe1843}  based on the records of daily observations between 1826 and 1843, discovered a frequency of occurrence of spots with a cycle to be approximately 10 years. Later Rudolph Wolf redefined the duration of a solar cycle finding  the maximum number of spots to be repeated every 11.1 years \citep{Wolf1852}. Wolf was the first researcher to organise permanent systematic observations of sunspots and introduced the concept of the daily “relative” sunspot number.

A significant step in improving the sunspot series was made in 1998 by \citet{hoyt1998a, hoyt1998b} who published a revised sunspot series with sunspot groups (GSN) from 1610 based on the analysis of 455242 records of 463 observers. However, not all the problems were solved in the comparison of sunspot records, which are often indistinct and were made in conditions of different visibility. 

The international team of researchers led by Leif Svalgaard has been trying to reconstruct an almost 400-year history of sunspot activity from 1610 to the 2000s \citep{Svalgaard2016, Svalgaard2017a, Svalgaard2017b}. The project was based on sketches by Christoph Scheiner, Johann Kaspar Staudacher, Heinrich Schwabe, Rudolf Wolf, and Hisako Koyama \citep{Carrasco2020, Hayakawa2020} considering two backbones: the Schwabe (1794 – 1883) and the Wolfer (1841 – 1944) backbone \citep{Svalgaard2016}.  

Although it appeared that the observations of sunspots was not consistent in different periods that restricted reliability of the built sunspot numbers for the period in the 18th century and in the first half of 19 century.  
Since July 2015, the SILSO International Data Center (Sunspot Index and Long-term Solar Observations) at the Royal Observatory of Belgium maintains a new, revised series of relative sunspot numbers (Version 2.0). New sunspot data have been considered  to be fairly reliable since 1750 \citep{Clette2014, Leussu2013, Svalgaard2016} and used by us for a further comparison.

A detailed analysis by \citet{Leussu2013} found  that  the reconstruction of \citet{hoyt1998b} on GSN series is more consistent if one uses of Schwabe’s data while the WSN records  of Wolf's sunspot reconstruction  is shown to decrease by roughly 20$\%$ around 1848. This is concluded to be caused by a change of the primary observer from Schwabe to Wolf. 

SS16 concluded  that it is clear that there was a dramatic change in the statistical properties of the data around 1810. Recent data revision shown the shifts of maxima of sunspot numbers to the cycles of early years (in 18 century) where originally were minima \citep{usoskin2021}. This highlights the fact that sunspot indices in the first 15 cycles could be build using not simply malign data but the data with many wrongly attributed times not known then to the person building the sunspot index,

Recently,  VH21  and VH2 explored the capacity of Bayesian models as a powerful tool to understand the complex nature of solar magnetic activity and to predict thhis solar activity for the next few cycles (VH21) as well as  extending their model for a description of the whole set of solar activity index defined by by sunspot numbers (VH22). The authors explored and re-emphasized the key distinctions in recent efforts to reproduce the sunspot numbers in the most verified fashion  based on the GSN by \citet{hoyt1998b} and SSN \citep{SILSO}. 

The authors also compared their data with the sunspot reconstructions by \citet{Clette2014}  and  \citet{Svalgaard2016} using a correction factor of 1.55 versus the correction factor of 1.208 in \citet{hoyt1998a, hoyt1998b}. 
Although, a comparison of the ML modelling results by VH21 and VH22  revealed that the fittings are dependent on the part of SSN data used in thhe training for Bayesian approach, e.g. the ML model fitting correctly either the Maunder minimum and current cycles (VH22) without any changes in cycles 25-27 or fitting the cycles of the 19th-21st century leading to prediction of the modern GSM (VH21), similarly to \citet{zhar15} but not recognising Maunder minimum. 

Despite large professional efforts of revising the existing sunspot index associated with sunspots there are still natural limitations how far these efforts can go in a lack of the data or known calendar dates. While keeping this index is a significant effort to link solar activity processes to the action of solar dynamo, the other options offered by the modern solar instruments providing solar magnetic field for the whole disk during the past 45 years are worth of investigation and consideration.

\citet{zhar15} suggested to use eigen vectors of solar background magnetic field derived with  Principal Component Analysis  from the synoptic maps obtained by Wilcox Solar Observatory to three (21-23) \citep{zharkova12} ,or four cycles 21-24 \citep{Zharkova2022} and derived the first eight significant eigen vectors covering the majority of the magnetic data by variance.     
The  summary curve of the two principal components (PCs) assigned to the magnetic waves generated by dipole magnetic sources, was suggested to be an additional solar activity index, whose modulus  curve is shown linked to the averaged sunspot numbers \citep{shepherd14, zhar15}.  This summary curve of two PCs was suggested as a new proxy of solar activity as it reflects the complementary entity of solar activity - poloidal magnetic field.

The summary curve calculated backward to 1200  and forward by 1200 years revealed very distinct variations of the cycle amplitudes in every 350-400 years, or grand solar cycles \citep{zhar15}. These grand solar cycles are separated by grand solar minima (GSMs),  when the amplitudes of 11 year cycles become very small, similar to those reported in Maunder,  Wolf and Oort  and other grand solar minima. It turned out that there is a steady decrease of the cycle amplitudes  from cycles 21 to 24 entering into the modern grand solar minimum (2020-2053).

In the current paper we attempted to establish the link for this new proxy, summary curve, or modulus summary curve, MSC, of the solar background magnetic field with the solar activity index defined by averaged sunspot numbers.  We made a comparison of MSC  with the whole set of sunspot cycle indices since 1700 and revealed a rather close correspondence of the cycle timings, duration and maxima times for the cycles fro 12 to 24, 6,7 and -4,-3. Although, there are some discrepancies, which occurred  between the maximum amplitudes, durations and shifts of maximum times  of MSC versus SSN curve in the specific intervals of 1720-1760 in the 18th century and of 1830-1860 in the 19th century. 

The most puzzling is the distribution of sunspot cycles in the 18th century  when some cycles had durations of 7 or 8 years while others lasted for 15 years. Contrary to this description, the MSC curve reveals pretty regular cycles with double maxima (cycles 1-4), again $\Psi$ type distribution between cycles 0 and 1 and for cycles -1 and -2 s before they come to Maunder minimum. The MSC cycles closer to 1700-1750  reveal smaller maximal magnitudes  in cycles -3 to 0 and in cycle 1-4 than the amplitudes of sunspot index, while cycles -2 to 0 have reversed maxima with minima between SSN and MSC curves.

These discrepancies can be partially explained  by either: a)  poor, or lack of, observations or b) a difference of the magnetic field entities  (poloidal for MSC versus toroidal for SSN magnetic field). The first reason can include  the absence of lead observers for sunspots, mixtures with the dates of sunspot drawings induced by changing from Julian to Gregorian calendars made at different times by different countries or even mix in years because of the different starts of New Years in  some counties. There is rather difficult to know, which of these reasons contribute, although, recent evaluation of solar cycle maxima times from the carbon 14 isotopes lent some support  to the MSC times of maxima as they were shifted in the MSC times direction.

The second reason is well understood in terms that the MSC is produced by poloidal magnetic field of the Sun while the averaged sunspot numbers are loosely associated  with toroidal magnetic field, though this link is rather distant via the number of sunspots in groups. They appearance differences  in a given grand solar cycle are clearly demonstrated by the simulations of these field by the two layer dynamo model with meridional circulations which successfully explained the observed summary curve and butterfly diagrams for  the  grand solar cycle and Maunder minimum. 

 Of course, potentially, it is rather logical to assume that PCA introduced some errors into the derived eigen vectors and their summary curve suggested as a new proxy for the solar activity index. However, the most recent investigation of eigen vectors of SBMF from cycles 21-24 \citep{Zharkova2022} has revealed that the addition of the magnetic field data  from extra solar cycle 24 did not change the eigen values and eigen vectors at least for the first four pair, or eight eigenvectors. This is assuring finding, which helps to consider the eigen vectors of poloidal field as a good proxy of solar activity, in addition to the existing solar indices defined by averaged sunspot numbers. 
 
 A usage of these two indices, SSN and MSC, or summary curve, which now become also available as an supplementary array for 2000 years, can pour a light into solar activity evolution from two different perspectives, toroidal magnetic field of sunspots and poloidal magnetic field of the solar background. This second proxy of SA will be a very essential addition for researchers in general, and during the modern grand solar minimum in cycles 25-27 (2020-2063), in particular.
\begin{acknowledgments}
The authors would like to thank the Solar Influences Data Analysis Center (SIDC) at the Royal Observatory of Belgium for providing the corrected averaged sunspot numbers. The authors also express their deepest gratitude to the staff and directorate of Wilcox Solar Observatory for providing the coherent long-term observations of full disk synoptic maps of the solar background magnetic field. 
\end{acknowledgments}
\bibliographystyle{aasjournal}
\bibliography{zhark_cycles}

\begin{thebibliography}{}
\expandafter\ifx\csname natexlab\endcsname\relax\def\natexlab#1{#1}\fi
\providecommand{\url}[1]{\href{#1}{#1}}
\providecommand{\dodoi}[1]{doi:~\href{http://doi.org/#1}{\nolinkurl{#1}}}
\providecommand{\doeprint}[1]{\href{http://ascl.net/#1}{\nolinkurl{http://ascl.net/#1}}}
\providecommand{\doarXiv}[1]{\href{https://arxiv.org/abs/#1}{\nolinkurl{https://arxiv.org/abs/#1}}}

\bibitem[{{Arlt}(2008)}]{Arlt2008}
{Arlt}, R. 2008, \solphys, 247, 399, \dodoi{10.1007/s11207-007-9113-4}

\bibitem[{{Arlt}(2009)}]{Arlt2009}
---. 2009, \solphys, 255, 143, \dodoi{10.1007/s11207-008-9306-5}

\bibitem[{{Arlt} {et~al.}(2016){Arlt}, {Senthamizh Pavai}, {Schmiel}, \&
  {Spada}}]{Arlt2016}
{Arlt}, R., {Senthamizh Pavai}, V., {Schmiel}, C., \& {Spada}, F. 2016, \aap,
  595, A104, \dodoi{10.1051/0004-6361/201629000}

\bibitem[{{Arlt} \& {Vaquero}(2020)}]{ArltVaquero2020}
{Arlt}, R., \& {Vaquero}, J.~M. 2020, Living Reviews in Solar Physics, 17, 1,
  \dodoi{10.1007/s41116-020-0023-y}

\bibitem[{{Baranyi} {et~al.}(2016){Baranyi}, {Gy{\H{o}}ri}, \&
  {Ludm{\'a}ny}}]{Baranyi2016}
{Baranyi}, T., {Gy{\H{o}}ri}, L., \& {Ludm{\'a}ny}, A. 2016, \solphys, 291,
  3081, \dodoi{10.1007/s11207-016-0930-1}

\bibitem[{{Beer} {et~al.}(1998){Beer}, {Tobias}, \& {Weiss}}]{Beer1998}
{Beer}, J., {Tobias}, S., \& {Weiss}, N. 1998, \solphys, 181, 237,
  \dodoi{10.1023/A:1005026001784}

\bibitem[{{Carrasco} {et~al.}(2020){Carrasco}, {Gallego}, \&
  {Vaquero}}]{Carrasco2020}
{Carrasco}, V.~M.~S., {Gallego}, M.~C., \& {Vaquero}, J.~M. 2020, \mnras, 496,
  2482, \dodoi{10.1093/mnras/staa1633}

\bibitem[{{Carrasco} {et~al.}(2021{\natexlab{a}}){Carrasco}, {Gallego},
  {Villalba {\'A}lvarez}, \& {Vaquero}}]{Carrasco2021a}
{Carrasco}, V.~M.~S., {Gallego}, M.~C., {Villalba {\'A}lvarez}, J., \&
  {Vaquero}, J.~M. 2021{\natexlab{a}}, \solphys, 296, 59,
  \dodoi{10.1007/s11207-021-01809-1}

\bibitem[{{Carrasco} {et~al.}(2021{\natexlab{b}}){Carrasco}, {Hayakawa},
  {Kuroyanagi}, {Gallego}, \& {Vaquero}}]{Carrasco2021b}
{Carrasco}, V.~M.~S., {Hayakawa}, H., {Kuroyanagi}, C., {Gallego}, M.~C., \&
  {Vaquero}, J.~M. 2021{\natexlab{b}}, \mnras, 504, 5199,
  \dodoi{10.1093/mnras/stab1155}

\bibitem[{{Carrasco} {et~al.}(2018){Carrasco}, {Vaquero}, {Trigo}, \&
  {Gallego}}]{Carrasco2018}
{Carrasco}, V.~M.~S., {Vaquero}, J.~M., {Trigo}, R.~M., \& {Gallego}, M.~C.
  2018, {A Curious History of Sunspot Penumbrae: An Update},
  \dodoi{10.1007/s11207-018-1328-z}

\bibitem[{{Chatzistergos} {et~al.}(2017){Chatzistergos}, {Usoskin},
  {Kovaltsov}, {Krivova}, \& {Solanki}}]{Chatzistergos2017}
{Chatzistergos}, T., {Usoskin}, I.~G., {Kovaltsov}, G.~A., {Krivova}, N.~A., \&
  {Solanki}, S.~K. 2017, \aap, 602, A69, \dodoi{10.1051/0004-6361/201630045}

\bibitem[{{Clette} {et~al.}(2014){Clette}, {Svalgaard}, {Vaquero}, \&
  {Cliver}}]{Clette2014}
{Clette}, F., {Svalgaard}, L., {Vaquero}, J.~M., \& {Cliver}, E.~W. 2014, \ssr,
  186, 35, \dodoi{10.1007/s11214-014-0074-2}

\bibitem[{{Cliver}(2016)}]{Cliver2016}
{Cliver}, E.~W. 2016, \solphys, 291, 2891, \dodoi{10.1007/s11207-016-0929-7}

\bibitem[{{Dreyer}(1903)}]{Dreyer1903}
{Dreyer}, J.~L.~E. 1903, The Observatory, 26, 461

\bibitem[{{Eddy}(1976)}]{Eddy1976}
{Eddy}, J.~A. 1976, Science, 192, 1189, \dodoi{10.1126/science.192.4245.1189}

\bibitem[{{Gy{\H{o}}ri} {et~al.}(2017){Gy{\H{o}}ri}, {Ludm{\'a}ny}, \&
  {Baranyi}}]{Gyori2017}
{Gy{\H{o}}ri}, L., {Ludm{\'a}ny}, A., \& {Baranyi}, T. 2017, \mnras, 465, 1259,
  \dodoi{10.1093/mnras/stw2667}

\bibitem[{{Hathaway}(2013)}]{Hathaway2013}
{Hathaway}, D.~H. 2013, \solphys, 286, 347, \dodoi{10.1007/s11207-013-0291-y}

\bibitem[{{Hathaway}(2015)}]{Hathaway2015}
---. 2015, Living Reviews in Solar Physics, 12, 4, \dodoi{10.1007/lrsp-2015-4}

\bibitem[{{Hathaway} {et~al.}(2002){Hathaway}, {Wilson}, \&
  {Reichmann}}]{Hathaway02}
{Hathaway}, D.~H., {Wilson}, R.~M., \& {Reichmann}, E.~J. 2002, \solphys, 211,
  357

\bibitem[{{Hayakawa} {et~al.}(2020){Hayakawa}, {Clette}, {Horaguchi}, {Iju},
  {Knipp}, {Liu}, \& {Nakajima}}]{Hayakawa2020}
{Hayakawa}, H., {Clette}, F., {Horaguchi}, T., {et~al.} 2020, \mnras, 492,
  4513, \dodoi{10.1093/mnras/stz3345}

\bibitem[{{Hayakawa} {et~al.}(2021{\natexlab{a}}){Hayakawa}, {Iju}, {Uneme},
  {Besser}, {Kosaka}, \& {Imada}}]{Hayakawa2021a}
{Hayakawa}, H., {Iju}, T., {Uneme}, S., {et~al.} 2021{\natexlab{a}}, \mnras,
  506, 650, \dodoi{10.1093/mnras/staa2965}

\bibitem[{{Hayakawa} {et~al.}(2017{\natexlab{a}}){Hayakawa}, {Iwahashi},
  {Tamazawa}, {Ebihara}, {Kawamura}, {Isobe}, {Namiki}, \&
  {Shibata}}]{Hayakawa2017b}
{Hayakawa}, H., {Iwahashi}, K., {Tamazawa}, H., {et~al.} 2017{\natexlab{a}},
  \pasj, 69, 86, \dodoi{10.1093/pasj/psx087}

\bibitem[{{Hayakawa} {et~al.}(2021{\natexlab{b}}){Hayakawa}, {Kuroyanagi},
  {Carrasco}, {Uneme}, {Besser}, {S{\^o}ma}, \& {Imada}}]{Hayakawa2021b}
{Hayakawa}, H., {Kuroyanagi}, C., {Carrasco}, V. M.~S., {et~al.}
  2021{\natexlab{b}}, \apj, 909, 166, \dodoi{10.3847/1538-4357/abd949}

\bibitem[{{Hayakawa} {et~al.}(2017{\natexlab{b}}){Hayakawa}, {Tamazawa},
  {Ebihara}, {Miyahara}, {Kawamura}, {Aoyama}, \& {Isobe}}]{Hayakawa2017a}
{Hayakawa}, H., {Tamazawa}, H., {Ebihara}, Y., {et~al.} 2017{\natexlab{b}},
  \pasj, 69, 65, \dodoi{10.1093/pasj/psx045}

\bibitem[{{Hoyt} \& {Schatten}(1998{\natexlab{a}})}]{hoyt1998b}
{Hoyt}, D.~V., \& {Schatten}, K.~H. 1998{\natexlab{a}}, \solphys, 181, 491,
  \dodoi{10.1023/A:1005056326158}

\bibitem[{{Hoyt} \& {Schatten}(1998{\natexlab{b}})}]{hoyt1998a}
---. 1998{\natexlab{b}}, \solphys, 179, 189, \dodoi{10.1023/A:1005007527816}

\bibitem[{{Hoyt} {et~al.}(1994){Hoyt}, {Schatten}, \& {Nesme-Ribes}}]{Hoyt1994}
{Hoyt}, D.~V., {Schatten}, K.~H., \& {Nesme-Ribes}, E. 1994, \grl, 21, 2067,
  \dodoi{10.1029/94GL01698}

\bibitem[{{Karoff} {et~al.}(2019){Karoff}, {J{\o}rgensen}, {Senthamizh Pavai},
  \& {Arlt}}]{Karoff2019}
{Karoff}, C., {J{\o}rgensen}, C.~S., {Senthamizh Pavai}, V., \& {Arlt}, R.
  2019, \solphys, 294, 78, \dodoi{10.1007/s11207-019-1466-y}

\bibitem[{{Kitiashvili}(2020)}]{Kitiashvili2020}
{Kitiashvili}, I.~N. 2020, \apj, 890, 36, \dodoi{10.3847/1538-4357/ab64e7}

\bibitem[{{Leussu} {et~al.}(2013){Leussu}, {Usoskin}, {Arlt}, \&
  {Mursula}}]{Leussu2013}
{Leussu}, R., {Usoskin}, I.~G., {Arlt}, R., \& {Mursula}, K. 2013, \aap, 559,
  A28, \dodoi{10.1051/0004-6361/201322373}

\bibitem[{{Livingston} {et~al.}(2012){Livingston}, {Penn}, \&
  {Svalgaard}}]{Livingston2012}
{Livingston}, W., {Penn}, M.~J., \& {Svalgaard}, L. 2012, \apjl, 757, L8,
  \dodoi{10.1088/2041-8205/757/1/L8}

\bibitem[{{Lockwood} {et~al.}(2014){Lockwood}, {Owens}, \&
  {Barnard}}]{Lockwood2014}
{Lockwood}, M., {Owens}, M.~J., \& {Barnard}, L. 2014, Journal of Geophysical
  Research (Space Physics), 119, 5183, \dodoi{10.1002/2014JA019972}

\bibitem[{{Lockwood} {et~al.}(2016){Lockwood}, {Owens}, \&
  {Barnard}}]{Lockwood2016b}
---. 2016, \solphys, 291, 2843, \dodoi{10.1007/s11207-016-0967-1}

\bibitem[{{Maunder}(1922)}]{Maunder1922}
{Maunder}, E.~W. 1922, Journal of the British Astronomical Association, 32, 140

\bibitem[{{Mu{\~n}oz-Jaramillo} \& {Vaquero}(2019)}]{Munoz-Jaramillo2019}
{Mu{\~n}oz-Jaramillo}, A., \& {Vaquero}, J.~M. 2019, Nature Astronomy, 3, 205,
  \dodoi{10.1038/s41550-018-0638-2}

\bibitem[{{Nagovitsyn} {et~al.}(2012){Nagovitsyn}, {Pevtsov}, \&
  {Livingston}}]{Nagovitsyn2012}
{Nagovitsyn}, Y.~A., {Pevtsov}, A.~A., \& {Livingston}, W.~C. 2012, \apjl, 758,
  L20, \dodoi{10.1088/2041-8205/758/1/L20}

\bibitem[{{Neuh{\"a}user} {et~al.}(2018){Neuh{\"a}user}, {Arlt}, \&
  {Richter}}]{Neuhauser2018}
{Neuh{\"a}user}, R., {Arlt}, R., \& {Richter}, S. 2018, Astronomische
  Nachrichten, 339, 219, \dodoi{10.1002/asna.201813481}

\bibitem[{{Obridko} {et~al.}(2021){Obridko}, {Sokoloff}, {Pipin}, {Shibalvaa},
  \& {Livshits}}]{Obridko2021}
{Obridko}, V.~N., {Sokoloff}, D.~D., {Pipin}, V.~V., {Shibalvaa}, A.~S., \&
  {Livshits}, I.~M. 2021, Monthly Notices of Royal Astr. Soc, 4990.
\newblock \doarXiv{2108.10527}

\bibitem[{{Ogurtsov}(2013)}]{Ogurtsov2013}
{Ogurtsov}, M.~G. 2013, Geomagnetism and Aeronomy, 53, 663,
  \dodoi{10.1134/S0016793213050137}

\bibitem[{{Parker}(1955)}]{Parker55}
{Parker}, E.~N. 1955, \apj, 122, 293, \dodoi{10.1086/146087}

\bibitem[{{Popova} {et~al.}(2013){Popova}, {Zharkova}, \& {Zharkov}}]{popova13}
{Popova}, E., {Zharkova}, V., \& {Zharkov}, S. 2013, Annales Geophysicae, 31,
  2023, \dodoi{10.5194/angeo-31-2023-2013}

\bibitem[{{Ribes} \& {Nesme-Ribes}(1993)}]{Ribes1993}
{Ribes}, J.~C., \& {Nesme-Ribes}, E. 1993, \aap, 276, 549

\bibitem[{{Schmidt} \& {Lipson}(2009)}]{Schmidt09}
{Schmidt}, M., \& {Lipson}, H. 2009, Science, 324, 81,
  \dodoi{10.1126/science.1165893}

\bibitem[{{Schove}(1983)}]{Schove1983}
{Schove}, D.~J. 1983, {Sunspot cycles.}

\bibitem[{{Schwabe}(1843)}]{Schwabe1843}
{Schwabe}, M. 1843, Astronomische Nachrichten, 20, 283,
  \dodoi{10.1002/asna.18430201706}

\bibitem[{{Senthamizh Pavai} {et~al.}(2015){Senthamizh Pavai}, {Arlt},
  {Dasi-Espuig}, {Krivova}, \& {Solanki}}]{Senthamizh2015}
{Senthamizh Pavai}, V., {Arlt}, R., {Dasi-Espuig}, M., {Krivova}, N.~A., \&
  {Solanki}, S.~K. 2015, \aap, 584, A73, \dodoi{10.1051/0004-6361/201527080}

\bibitem[{{Shepherd} {et~al.}(2014){Shepherd}, {Zharkov}, \&
  {Zharkova}}]{shepherd14}
{Shepherd}, S.~J., {Zharkov}, S.~I., \& {Zharkova}, V.~V. 2014, \apj, 795, 46,
  \dodoi{10.1088/0004-637X/795/1/46}

\bibitem[{{{SILSO} World Data Center}(2021)}]{SILSO}
{{SILSO} World Data Center}. 2021, International Sunspot Number Monthly
  Bulletin and online catalogue, $https://wwwbis.sidc.be/silso/datafiles$

\bibitem[{{Simpson}(2020)}]{Simpson2020}
{Simpson}, J. 2020, Journal of the British Astronomical Association, 130, 15

\bibitem[{{Solanki} {et~al.}(2004){Solanki}, {Usoskin}, {Kromer},
  {Sch{\"u}ssler}, \& {Beer}}]{solanki04}
{Solanki}, S.~K., {Usoskin}, I.~G., {Kromer}, B., {Sch{\"u}ssler}, M., \&
  {Beer}, J. 2004, \nat, 431, 1084, \dodoi{10.1038/nature02995}

\bibitem[{{Soon} \& {Yaskell}(2003)}]{SoonYaskell2003}
{Soon}, W. W.-H., \& {Yaskell}, S.~H. 2003, {The Maunder Minimum and the
  Variable Sun-Earth Connection}, \dodoi{10.1142/5199}

\bibitem[{{Stix}(1976)}]{Stix76}
{Stix}, M. 1976, \aap, 47, 243

\bibitem[{Suykens {et~al.}(2002)Suykens, Van~Gestel, De~Brabanter, De~Moor, \&
  Vandewalle}]{suykens2002}
Suykens, J.~A., Van~Gestel, T., De~Brabanter, J., De~Moor, B., \& Vandewalle,
  J.~P. 2002, Least squares support vector machines (World scientific)

\bibitem[{{Svalgaard}(2017)}]{Svalgaard2017}
{Svalgaard}, L. 2017, \solphys, 292, 4, \dodoi{10.1007/s11207-016-1023-x}

\bibitem[{{Svalgaard} \& {Schatten}(2016)}]{Svalgaard2016}
{Svalgaard}, L., \& {Schatten}, K.~H. 2016, \solphys, 291, 2653,
  \dodoi{10.1007/s11207-015-0815-8}

\bibitem[{{Svalgaard} \& {Schatten}(2017{\natexlab{a}})}]{Svalgaard2017b}
---. 2017{\natexlab{a}}, arXiv e-prints, arXiv:1705.02024.
\newblock \doarXiv{1705.02024}

\bibitem[{{Svalgaard} \& {Schatten}(2017{\natexlab{b}})}]{Svalgaard2017c}
---. 2017{\natexlab{b}}, arXiv e-prints, arXiv:1706.01154.
\newblock \doarXiv{1706.01154}

\bibitem[{{Svalgaard} \& {Schatten}(2017{\natexlab{c}})}]{Svalgaard2017a}
---. 2017{\natexlab{c}}, arXiv e-prints, arXiv:1704.07061.
\newblock \doarXiv{1704.07061}

\bibitem[{{Tamazawa} {et~al.}(2017){Tamazawa}, {Kawamura}, {Hayakawa},
  {Tsukamoto}, {Isobe}, \& {Ebihara}}]{Tamazawa2017}
{Tamazawa}, H., {Kawamura}, A.~D., {Hayakawa}, H., {et~al.} 2017, \pasj, 69,
  22, \dodoi{10.1093/pasj/psw132}

\bibitem[{{Usoskin}(2017)}]{Usoskin2017}
{Usoskin}, I.~G. 2017, Living Reviews in Solar Physics, 14, 3,
  \dodoi{10.1007/s41116-017-0006-9}

\bibitem[{{Usoskin} {et~al.}(2004){Usoskin}, {Mursula}, {Solanki},
  {Sch{\"u}ssler}, \& {Alanko}}]{usoskin2004}
{Usoskin}, I.~G., {Mursula}, K., {Solanki}, S., {Sch{\"u}ssler}, M., \&
  {Alanko}, K. 2004, \aap, 413, 745, \dodoi{10.1051/0004-6361:20031533}

\bibitem[{{Usoskin} {et~al.}(2021){Usoskin}, {Solanki}, {Krivova}, {Hofer},
  {Kovaltsov}, {Wacker}, {Brehm}, \& {Kromer}}]{usoskin2021}
{Usoskin}, I.~G., {Solanki}, S.~K., {Krivova}, N.~A., {et~al.} 2021, \aap, 649,
  A141, \dodoi{10.1051/0004-6361/202140711}

\bibitem[{{Vaquero}(2007)}]{Vaquero2007b}
{Vaquero}, J.~M. 2007, Advances in Space Research, 40, 929,
  \dodoi{10.1016/j.asr.2007.01.087}

\bibitem[{{Vaquero} \& {Gallego}(2014)}]{Vaquero2014}
{Vaquero}, J.~M., \& {Gallego}, M.~C. 2014, Advances in Space Research, 53,
  1162, \dodoi{10.1016/j.asr.2014.01.015}

\bibitem[{{Vaquero} {et~al.}(2007){Vaquero}, {Gallego}, \&
  {Trigo}}]{Vaquero2007a}
{Vaquero}, J.~M., {Gallego}, M.~C., \& {Trigo}, R.~M. 2007, Advances in Space
  Research, 40, 1895, \dodoi{10.1016/j.asr.2007.02.097}

\bibitem[{{Vasiljeva} \& {Pishkalo}(2021)}]{Vasiljeva2021}
{Vasiljeva}, I.~E., \& {Pishkalo}, M.~I. 2021, Kinematics and Physics of
  Celestial Bodies, 37, 200, \dodoi{10.3103/S0884591321040073}

\bibitem[{{Velasco Herrera} {et~al.}(2022){Velasco Herrera}, {Soon}, {Hoyt}, \&
  {Murakzy}}]{velasco2022}
{Velasco Herrera}, V.~M., {Soon}, W., {Hoyt}, D.~V., \& {Murakzy}, J. 2022,
  Solar Phys., 297, 8, \dodoi{10.1007/s11207-021-01926-x}

\bibitem[{{Velasco Herrera} {et~al.}(2021){Velasco Herrera}, {Soon}, \&
  {Legates}}]{velasco2021}
{Velasco Herrera}, V.~M., {Soon}, W., \& {Legates}, D.~R. 2021, Advances in
  Space Research, 68, 1485, \dodoi{10.1016/j.asr.2021.03.023}

\bibitem[{{Vokhmyanin} {et~al.}(2021){Vokhmyanin}, {Arlt}, \&
  {Zolotova}}]{Vokhmyanin2021}
{Vokhmyanin}, M., {Arlt}, R., \& {Zolotova}, N. 2021, \solphys, 296, 4,
  \dodoi{10.1007/s11207-020-01752-7}

\bibitem[{{Waldmeier}(1961)}]{waldmeier61}
{Waldmeier}, M. 1961, {The sunspot-activity in the years 1610-1960}

\bibitem[{{Willamo} {et~al.}(2017){Willamo}, {Usoskin}, \&
  {Kovaltsov}}]{Willamo2017}
{Willamo}, T., {Usoskin}, I.~G., \& {Kovaltsov}, G.~A. 2017, \aap, 601, A109,
  \dodoi{10.1051/0004-6361/201629839}

\bibitem[{{Wolf}(1852)}]{Wolf1852}
{Wolf}, M. 1852, \mnras, 13, 29, \dodoi{10.1093/mnras/13.1.29}

\bibitem[{{Wolf}(1850{\natexlab{a}})}]{wolf1850a}
{Wolf}, R. 1850{\natexlab{a}}, Astronomische Mitteilungen der
  Eidgen\&ouml;ssischen Sternwarte Zurich, 1, 3

\bibitem[{{Wolf}(1850{\natexlab{b}})}]{wolf1850b}
---. 1850{\natexlab{b}}, Astronomische Mitteilungen der Eidgen\&ouml;ssischen
  Sternwarte Zurich, 1, 15

\bibitem[{{Wolf}(1877)}]{Wolf1877}
---. 1877, {Geschichte der astronomie}

\bibitem[{{Zharkov} {et~al.}(2008){Zharkov}, {Gavryuseva}, \&
  {Zharkova}}]{zharkov08}
{Zharkov}, S., {Gavryuseva}, E., \& {Zharkova}, V. 2008, \solphys, 248, 339,
  \dodoi{10.1007/s11207-007-9109-0}

\bibitem[{Zharkova(2020)}]{zhar2020}
Zharkova, V. 2020, Temperature, 7, 217, \dodoi{10.1080/23328940.2020.1796243}

\bibitem[{{Zharkova} {et~al.}(2018{\natexlab{a}}){Zharkova}, {Popova},
  {Shepherd}, \& {Zharkov}}]{Zharkova2018b}
{Zharkova}, V., {Popova}, E., {Shepherd}, S., \& {Zharkov}, S.
  2018{\natexlab{a}}, Journal of Atmospheric and Solar-Terrestrial Physics,
  176, 72, \dodoi{10.1016/j.jastp.2017.09.019}

\bibitem[{{Zharkova} \& {Shepherd}(2022)}]{Zharkova2022}
{Zharkova}, V.~V., \& {Shepherd}, S.~J. 2022, Monthly Notices of Royal
  Astron.Soc., in press, 17, \dodoi{10.1111/j.1365-2966.2012.21436.x}

\bibitem[{{Zharkova} {et~al.}(2015){Zharkova}, {Shepherd}, {Popova}, \&
  {Zharkov}}]{zhar15}
{Zharkova}, V.~V., {Shepherd}, S.~J., {Popova}, E., \& {Zharkov}, S.~I. 2015,
  Nature Scientific Reports, 5, 15689, \dodoi{doi:10.1038/srep15689}

\bibitem[{{Zharkova} {et~al.}(2018{\natexlab{b}}){Zharkova}, {Shepherd},
  {Popova}, \& {Zharkov}}]{Zharkova2018a}
{Zharkova}, V.~V., {Shepherd}, S.~J., {Popova}, E., \& {Zharkov}, S.~I.
  2018{\natexlab{b}}, in Space Weather of the Heliosphere: Processes and
  Forecasts, ed. C.~{Foullon} \& O.~E. {Malandraki}, Vol. 335, 211--215,
  \dodoi{10.1017/S1743921317010912}

\bibitem[{{Zharkova} {et~al.}(2012){Zharkova}, {Shepherd}, \&
  {Zharkov}}]{zharkova12}
{Zharkova}, V.~V., {Shepherd}, S.~J., \& {Zharkov}, S.~I. 2012, \mnras, 424,
  2943, \dodoi{10.1111/j.1365-2966.2012.21436.x}

\bibitem[{{Zito}(2016)}]{Zito2016}
{Zito}, R.~R. 2016, Sociology and Anthropology, 4, 953,
  \dodoi{10.13189/sa.2016.041102}

\end{thebibliography}
\end{document}